\documentclass{aa}  

\usepackage{graphicx}
\usepackage{txfonts}
\usepackage{lipsum}
\usepackage{subcaption}       
\usepackage{booktabs}
\usepackage{lscape}             
\usepackage{placeins}           
                                
\begin{document}


\title{Angular Momentum of Planet-Forming Disks: Implications for Infall Driven Misalignments}



\author{
Aashish Gupta\inst{1}\fnmsep\thanks{aashishgupta@virginia.edu}
\and
Cristiano Longarini\inst{2}\thanks{cl2000@cam.ac.uk}
\and
L. Ilsedore Cleeves\inst{1}
\and
Giovanni P. Rosotti\inst{3}
\and
Edwin A. Bergin\inst{4}
\and
Cathie J. Clarke\inst{2}
\and
Michael Küffmeier\inst{5}
\and
Zhi-Yun Li\inst{1}
}

\institute{
Department of Astronomy, University of Virginia, Charlottesville, VA 22904, USA
\and
Institute of Astronomy, University of Cambridge, Madingley Road, Cambridge CB3 0HA, UK
\and
Dipartimento di Fisica, Università degli Studi di Milano, Via Celoria 16, 20133 Milano, Italy
\and
Department of Astronomy, University of Michigan, Ann Arbor, MI 48109, USA
\and
Niels Bohr Institute, University of Copenhagen, Øster Voldgade 5, 1350 Copenhagen, Denmark
}



\abstract
{A significant fraction ($\gtrsim30\%$) of planet-forming disks and planetary systems are misaligned with respect to the rotational axis of their host stars, yet the dominant mechanism responsible for these misalignments remains unclear.}
{We aim to observationally constrain the angular momentum of Class II protoplanetary disks and assess whether late-stage infall of material can bring sufficient angular momentum to tilt them.} 
{We first computed the angular momenta of 15 disks with surface density profiles inferred from dynamical modeling of high angular resolution ALMA observations. Based on this sample, we derived a relation linking disk angular momentum to stellar mass, disk mass, and the radius enclosing 90\% of the $^{13}$CO flux and used it to estimate angular momenta of 18 more disks. We then compared disk values with theoretical predictions for late-stage accretion from clouds and observed streamers.}
{Angular momentum for most disks is lower than what theoretical models predict for late infall. This is also in qualitative agreement with comparison with streamer observations, however, characterization of mass of reservoirs feeding the streamers is needed to confirm this picture.}
{Interactions with nearby clouds, resulting in late-stage infall of material onto Class II disks, can potentially explain the observed misalignments within disks and planetary systems.}

\keywords{Planets and satellites: formation, Protoplanetary disks, Stars: formation}

\maketitle
\nolinenumbers

\section{Introduction} \label{sec:intro}

The classical view of star formation suggests that as a prestellar core collapses from $\sim10,000$~au to $\sim100$~au scales, most of its material is either accreted onto a central protostar or settles into a protoplanetary disk around it \citep[e.g.,][]{Terebey1984}. This system is typically assumed to evolve in isolation, with the protostar and disk rotating in roughly the same direction \citep[e.g.,][]{Bodenheimer1995}. Consequently, planets forming within these disks are expected to inherit the same orbital orientation around their host star.

However, recent observations challenge this traditional picture.
Measurements of stellar obliquities, i.e., the angles between a planet’s orbital axis and its host star's rotational axis, reveal that roughly half of all observed planetary systems are significantly misaligned with respect to their stars \citep[e.g.][]{Albrecht2022,Bowler2023}. 
Likewise, \cite{Biddle2025} reported that about $\gtrsim30\%$ of planet-forming disks are tilted relative to their central stars, while \cite{Bohn2022} found a similar fraction ($\gtrsim30\%$) exhibiting misalignments between the innermost ($<1$~au) and outermost ($>10$~au) regions of disks. Moreover, analysis of disk orientations around dipper stars \citep{Ansdell2020}, observed shadows in near-infrared scattered light observations of disks \citep[e.g.,][]{Ginski2021,Benisty2023,Garufi2024} and signatures of warps within disk \citep[][]{Winter2025, Kimmig2025} further suggest that misalignments between inner and outer disks are common.
Even within our Solar System, the Sun’s rotation axis is tilted by about $\sim6^{\circ}$ relative to the orbital plane of the planets \citep[e.g.,][]{Adams2010}.
Taken together, these studies suggest that processes capable of tilting protoplanetary disks are common, and that these misalignments are likely imprinted onto the planetary systems they form.

Although protoplanetary disks are traditionally assumed to be isolated systems, observationally we know that they are generally within a group of young stars embedded in large-scale ($\gtrsim1$~pc) molecular clouds \citep[e.g.,][]{Luhman2020,Gupta2022}. Consequently, these disks can interact with the nearby stars \citep[e.g.,][]{Cuello2023} and with clumps of molecular gas, sometimes referred to as `cloudlets' \citep[e.g.,][]{Dullemond2019}. Misalignments arising from stellar interactions are expected to be small and short-lived \citep[e.g.,][]{Nealon2020}. On the other hand, several theoretical \cite[e.g.,][]{Padoan2025}, numerical \citep[e.g.,][]{Thies2011,Kuffmeier2021,Kuffmeier2024,Pelkonen2025}, and observational \citep[e.g.,][]{Ginski2021,Tanious2024,Gupta2026} studies show that interaction with cloudlets, resulting in infalling streamers, can induce significant misalignments which can persist for longer if infall is sustained. This is primarily because the material falling at later stages is dynamically unrelated to the original core, and thus, have a completely different orientation of angular momentum.

However, to understand how easy it is to misalign the disk via infall, we need to compare the angular momentum of disks and infalling material. 
For example, the numerical simulations presented in \citet{Kuffmeier2021} show that if angular momentum of infalling material is comparable to that of disks, the resulting re-orientation of the disks can be quite long lasting. 
Despite angular momentum being a fundamental property of disks and being central to understanding disk formation and evolution (further discussed in Appendix \ref{app:other}), it is observationally poorly constrained. This is primarily because the distribution of mass along the disks, i.e., their surface density, is hard to constrain. However, detailed modeling of rotational velocities of disks can allow us to infer their surface denisities, as has been done for some of the biggest and brightest disks in nearby star-forming regions \citep[][]{Martire2024,Longarini2025}.

Here, we use the inferred surface density profiles to directly estimate angular momentum for 15 disks and derive an empirical relationship to estimate angular momentum of more representative disks, as described in Section \ref{sec:disks}. We also compiled observational and theoretical estimates on angular momentum of infalling material, as described in Section \ref{sec:infall}. Their comparisons are presented in Section \ref{sec:results} and our key findings are summarized in Section \ref{sec:summary}.

\section{Angular momentum of disks} \label{sec:disks}

\subsection{Sample}  \label{sec:sample}

For this study, we directly derive angular momentum of disks with surface density profiles inferred from high spatial ($\sim20$~au) and spectral ($\sim0.1$~km~s$^{-1}$) resolution data observed within ALMA Large Programs: MAPS \citep[project code: 2018.1.01055.L;][]{Oberg2021} and exoALMA \citep[project code: 2021.1.01123.L;][]{Teague2025}.
These studies targeted the biggest disks in the nearby ($\lesssim 160$~pc) star-forming region. 
As discussed in Section \ref{sec:ang_mom}, based on the above sample, we also derive a relation between angular momentum of disks and more typically estimated system properties such as disk mass ($M_{\rm d}$), stellar mass ($M_{*}$), and disk radii enclosing 90$\%$ of $^{13}$CO flux (R$_{\rm disk; 13CO}$). We apply the derived relation to further estimate angular momentum of 18 more typical disks observed within the ALMA Large Program AGEPRO \citep[project code: 2021.1.00128.L;][]{Zhang2025}. The complete sample of disks analyzed in this study is presented in Table \ref{tab:disk_properties}.

\subsection{Surface density}  \label{sec:surface_density}

In \citet{Longarini2025}, the surface density profiles of protoplanetary disks are inferred through dynamical modeling of rotation curves derived from exoALMA CO (3--2) and $^{13}$CO (3--2) observations \citep{Stadler2025}, using \textsc{DYSC}\footnote{https://github.com/crislong/DySc} code.
Similar procedure was applied by \citet{Martire2024} on MAPS CO (2--1) and $^{13}$CO (2--1) observations.

\begin{figure}[htbp]
    \centering
    \begin{subfigure}{0.48\textwidth}
        \centering
        \includegraphics[width=0.9\linewidth]{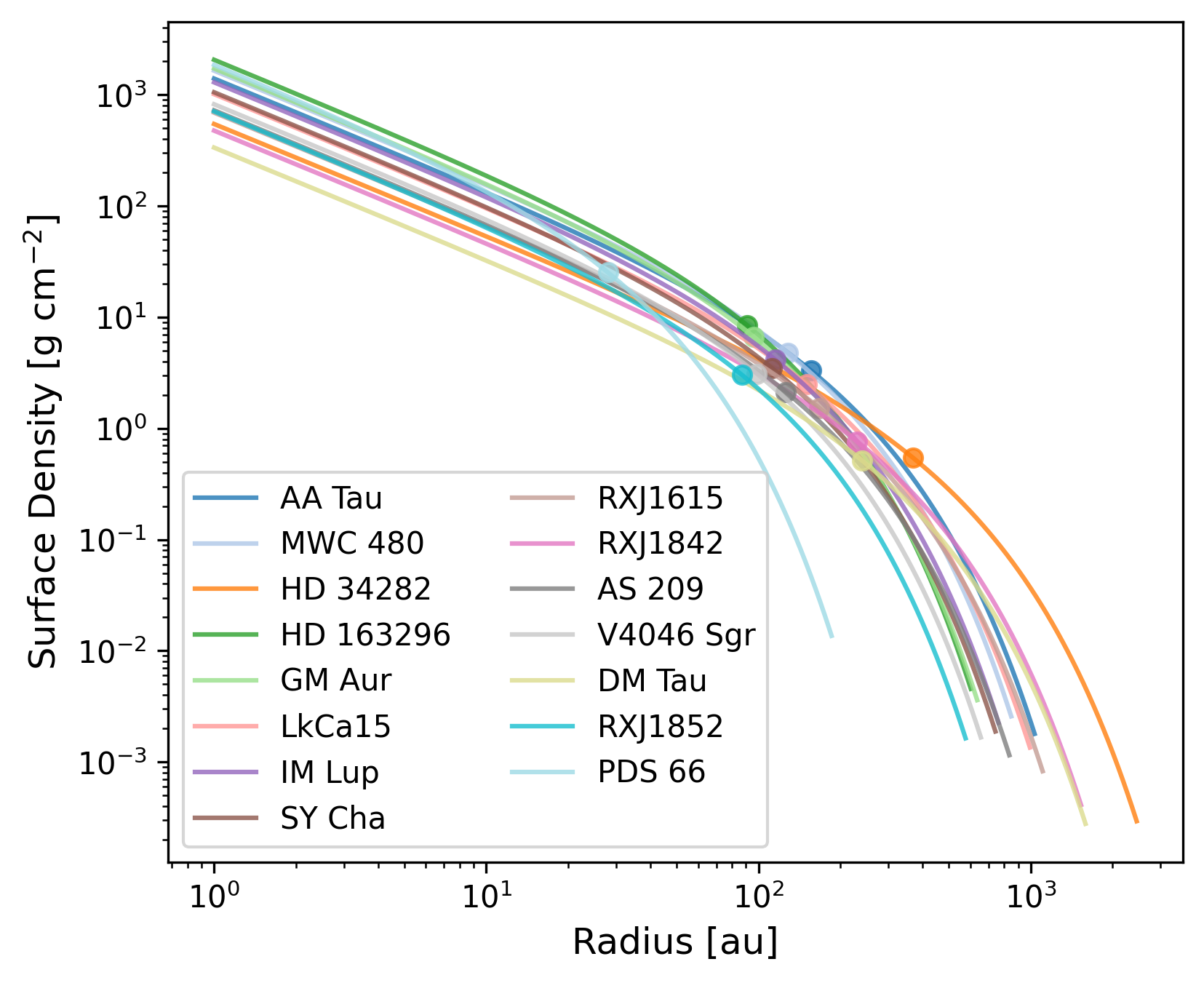}
        \label{fig:SDenProfile}
    \end{subfigure}
    \begin{subfigure}{0.48\textwidth}
        \centering
        \includegraphics[width=0.9\linewidth]{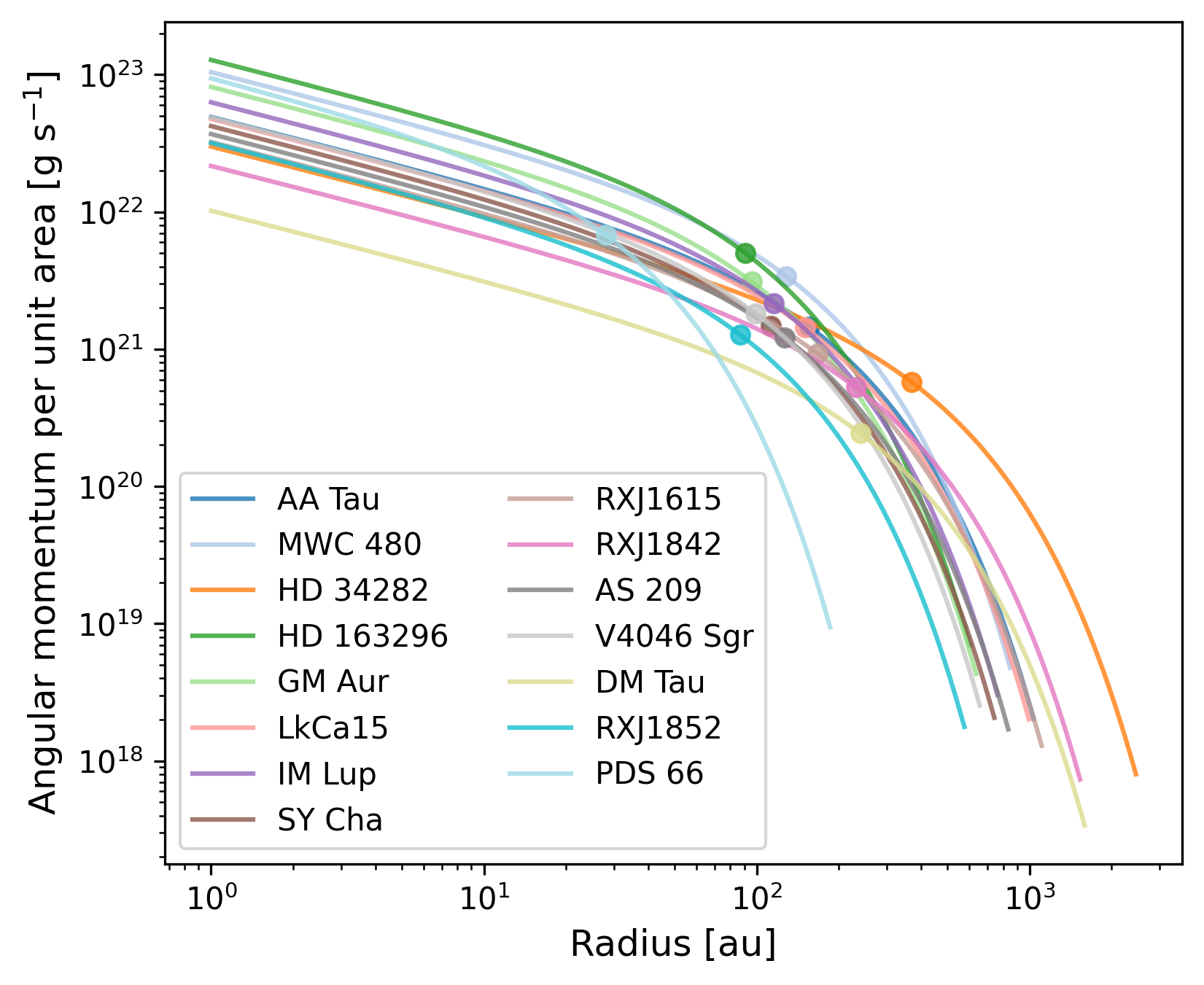}
        \label{fig:AngMomProfile}
    \end{subfigure}
    \caption{Top panel: surface density profiles of disks observed within exoALMA and MAPS large programs, as reported in \citet{Longarini2025} and \citet{Martire2024}, respectively. Bottom panel: angular momentum profiles of these disks as inferred by multiplying surface density profiles with Keplerian rotation. The circles on both the profiles represent the location of scale radii ($R_c$).}
    \label{fig:profiles}
\end{figure}

Both studies use a thermally stratified disk model, where the temperature and density structures are in hydrostatic equilibrium.
The rotational model employed by \citet{Martire2024} and \citet{Longarini2025} accounts for deviations from pure Keplerian rotation due to disk thickness, pressure gradients, and self-gravity.
This enables direct constraints on stellar mass ($M_\ast$), disk mass ($M_{\rm disk}$), and and scale radius of the disk ($R_{\rm c}$). Together, these allow to constrain the surface density profile ($\Sigma(R)$) for each disk.
In both studies, the surface density is assumed to follow a self-similar solution, as suggested in 
\citet{LyndenBell1974} :
\begin{equation} \label{equ:sdenity}
    \Sigma(R) = \frac{(2-\gamma)M_{\rm disk}}{2\pi R_{\rm c}^2}
    \left(\frac{R}{R_{\rm c}}\right)^{-\gamma}
    \exp\left[-\left(\frac{R}{R_{\rm c}}\right)^{2-\gamma}\right]  
\end{equation}
where $\gamma$ the surface-density slope. By default, both studies derived surface densities assuming $\gamma=1$. We discuss the impact of variations $\gamma$ in Appendix \ref{app:gamma} and for most cases, the changes in derived disk properties are within the intrinsic uncertainties. The surface density profiles, and corresponding angular momentum profiles, are shown in Figure \ref{fig:profiles}.

\subsection{Angular momentum} \label{sec:ang_mom}

Angular momentum of a ring of material, at a distance of $R$ from the star, with infinitesimally small mass ($dM$) and width ($dR$) can be represented as $dL = dM R^2 \omega$, where $\omega$ is the angular velocity. As the azimuthal velocity field is still strongly dominated by Keplerian rotation, we can use $\omega = \sqrt{GM_{\rm *}/R^{3}}$, where $G$ is the universal constant of gravity. Furthermore, as we know the surface density of the disk from Equation \ref{equ:sdenity}, the mass of the disk can be represented as $dM=\Sigma(R)2\pi R dR$.
The total angular momentum of the disk ($L$) can then be inferred as:
\begin{equation}
    L = 2\pi\sqrt{GM_{\rm *}}\int_{0}^{R_{\rm disk}}\Sigma(R)R^{1.5}dR
\end{equation}
where $R_{\rm disk}$ represents the total disk radius. It is not straight forward to define a boundary where the disk ends. For this work, we assumed $R_{\rm disk}=6.63 R_{\rm c}$, which is theoretically expected to enclose the $99\%$ mass of the disk. 
We note that slight variations in $R_{\rm disk}$ do not significantly affect our results, as the surface density drops exponentially in the outer disk regions, contributing negligibly to the total angular momentum (see Figure \ref{fig:profiles}). For example, halving $R_{\rm disk}$ reduces the resulting angular momentum only by $\sim8\%$.

Interestingly, we found the angular momentum of these disks to be strongly correlated with the disk radii enclosing 90$\%$ of $^{13}$CO flux (R$_{\rm disk; 13CO}$) (Spearman correlation coefficient: 0.59, p-value: 0.02). For comparison, the correlation with radii enclosing 90$\%$ of CO flux was much weaker, with Spearman correlation coeff. of 0.36 and corresponding p-value of 0.19. Both sets of radii are compared to angular momentum values in Appendix \ref{app:radii}. 

\begin{figure}[htbp]
\centering
\includegraphics[width=0.98\linewidth]{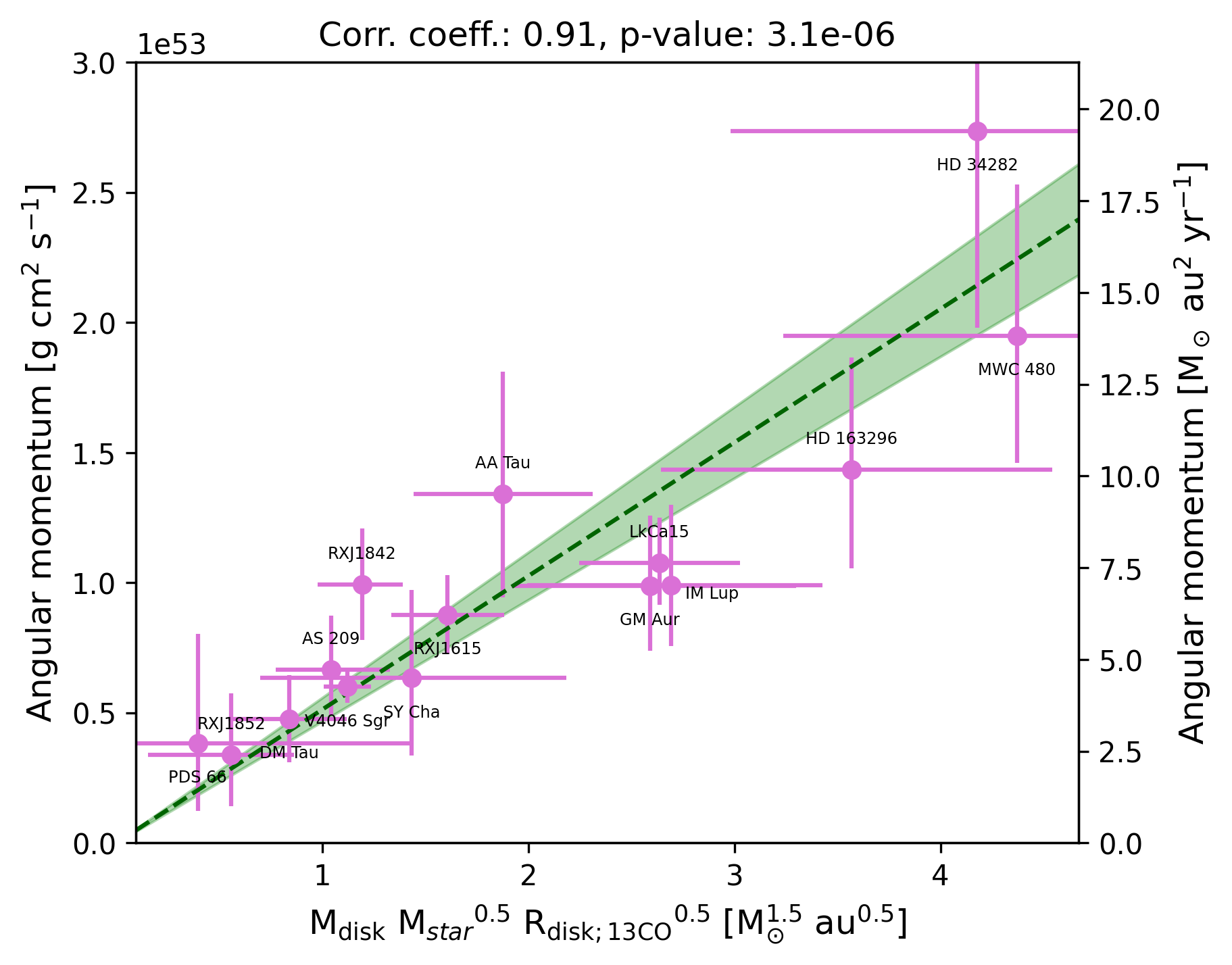}
\caption{Physically motivated relationship to derive angular momentum of disks using disk mass ($M_{\rm disk}$), stellar mass ($M_{*}$), and radius enclosing 90\% of the $^{13}$CO flux ($R_{\rm disk; 13CO}$), as discussed in Section \ref{sec:ang_mom}. The purple circles denote disks with observationally estimated surface density profiles, as discussed in Section \ref{sec:sample}. The green line and shaded region represents the derived relation and associated uncertainty (Equation \ref{equ:DiskAngMomRel}).}
\label{fig:AngMomEst}
\end{figure}

The significant correlation with R$_{\rm disk; 13CO}$ may suggest that it is proportional to the radius of gyration ($R_{\rm gyration}$) of disks, i.e., the radius at which disk mass can be considered concentrated without changing its moment of inertia. 
Therefore, by definition: 
\begin{equation}
    L = \sqrt{GM_{*}}*M_{\rm disk}*R_{\rm gyration}^{0.5}
\end{equation}
where $M_{\rm disk}$ denotes the disk mass.
Then we expect the disk angular momentum ($L$) to be related to R$_{\rm disk; 13CO}$ as:
\begin{equation}
    L \propto M_{*}^{0.5} M_{\rm disk} R_{\rm disk; 13CO}^{0.5}
\end{equation}
Indeed we find a stronger correlation (Spearman correlation coefficient: 0.91, p-value: $3.1\times10^{-6}$) between the two quantities, as shown in Figure \ref{fig:AngMomEst}. 
We fit this relation by minimizing $\chi^{2}$ deviation using \texttt{scipy.optimize.minimize\_scalar}. This allowed us to include asymmetric errors in both the independent and dependent variables (see Section \ref{sec:disk_errors}). The best-fit relation, as shown in Figure \ref{fig:AngMomEst}, is
\begin{equation} \label{equ:DiskAngMomRel}
\begin{aligned}
L ={}&(5.1\times10^{52} \pm 4.6\times10^{51}) \\
&\times \left(\frac{M_{*}}{M_{\odot}}\right)^{0.5}
\left(\frac{M_{\rm disk}}{M_{\odot}}\right)
\left(\frac{R_{\rm disk; 13CO}}{\mathrm{AU}}\right)^{0.5}
\,\mathrm{g\,cm^{2}\,s^{-1}}
\end{aligned}
\end{equation}

Using Equation \ref{equ:DiskAngMomRel}, we can estimate angular momentum of more typical disks, which are more compact making it harder to directly infer their surface density profiles. ALMA Large Program AGEPRO \citep[][]{Zhang2025} observed ten Class II disks each in Lupus \citep[][]{Deng2025} and Upper Scorpius \citep[][]{AgurtoGangas2025} star-forming regions. For these disks, the gas masses were estimated by simultaneously fitting emission from CO isotopologues and N$_2$H$^+$, a molecule produced when CO freezes out, as described in \citet{Trapman2025} and tabulated here in Table \ref{tab:disk_properties}. 

Within this program, the disks were also observed in $^{13}$CO (2--1) emission line, and the radial profiles for eighteen disks are publicly available \footnote{https://agepro.das.uchile.cl/data\_page} \citep[also see][]{Deng2025, AgurtoGangas2025}. We used these radial profiles to derive $R_{\rm disk; 13CO}$ for these disks. For this, we integrated total disk flux along the radial profile, till the flux at the given radius becomes equal to or less than zero. Then we select the radius within which the integrated flux was equal to 0.9 times the total disk flux. 
Combining the reported stellar and disk masses \citep{Deng2025,AgurtoGangas2025,Trapman2025} with our derived $R_{\rm disk; 13CO}$, we estimated the angular momentum of these disks, as listed in Table \ref{tab:disk_properties}.

A key assumption in applying Equation~\ref{equ:DiskAngMomRel} to estimate the angular momentum of smaller AGEPRO disks is that they have surface density profiles structurally similar to the larger disks. In other words we assume the small disks to be scaled down versions of bigger disks. This assumption is theoretically justified by the scale-invariance property of the \citet{LyndenBell1974} solutions, used to infer surface density profiles (Section \ref{sec:surface_density}). However, it is not straightforward to directly constrain surface density profiles of small disks, needed to test this assumption. Observationally, \citet{Trapman2025} reported a strong relation between $M_{\rm disk}$ and the size of disk as measured using CO (2--1). This relationship persisted for roughly four orders of magnitude in $M_{\rm disk}$ and two orders of magnitude in disk size, where the most massive disks were comparable to the disks used to derive the angular momentum relationship (see middle panel in Figure \ref{fig:InfallComparison}). Similar relationships have also been observed in larger samples using dust continuum observations of disks \citep[e.g.,][]{Andrews2018}. This suggests that our assumption of small disks being scaled down versions of big disks is generally valid.

\subsection{Uncertainties} \label{sec:disk_errors}

For disks with surface density profiles estimated in \citet{Longarini2025}, uncertainties on the angular momentum were estimated via Monte Carlo error propagation, where input parameters ($M_{*}$, $M_{\rm disk}$, $R_c$) were sampled from their reported asymmetric error distributions and uncertainties defined by the 16th and 84th percentiles. The uncertainties on the surface density parameters in \citet{Longarini2025} come from two primary sources. First, there is an error associated with the rotation curve extraction, which is given as an input for the MCMC fitting procedure. This error represents a standard deviation derived from the azimuthal averaging of the rotational maps, following the methodology described in \citet{Izquierdo2025}. Second, the uncertainties on the thermal structure are propagated by running the fitting procedure 100 times, each time sampling the thermal structure parameters from the posterior distribution of \citet{GallowaySprietsma2025}. The outputs from all walkers are combined to construct the final posterior distributions, from which the formal error is estimated using a Gaussian confidence interval.

The errors reported in \citet{Martire2024} are likely underestimated, as the uncertainties on the thermal structure were not propagated to the disk surface density parameters. For these disks, we therefore assumed relative uncertainties in all input parameters similar to those derived in \citet{Longarini2025} ($\sim2\%$ in $M_{*}$, $\sim20$--$25\%$ in $M_{\rm d}$ and $R_{\rm c}$), and propagated them to the angular momentum values using the same Monte Carlo approach.

For disks observed as part of the AGEPRO large program \citep{Deng2025,AgurtoGangas2025}, we expect the uncertainties in angular momentum to be dominated by uncertainties in disk masses \citep{Trapman2025}, together with the uncertainty in the scaling relation in Equation~\ref{equ:DiskAngMomRel}. These uncertainties were propagated to the angular momentum values using similar Monte Carlo error propagation.

\section{Angular momentum of infalling material} \label{sec:infall}

\subsection{Theoretical predictions} \label{sec:infall_pred}

To understand how likely it is for infalling material to tilt disks, we want to understand how the angular momentum of infalling material compares to the disks', as derived in Section \ref{sec:disks}. Theoretically, the infall of material onto a point source of gravity (a Class II star) from surrounding gas (molecular clouds) can approximated as Bond-Hoyle accretion \citep[e.g.,][]{Bondi1944,Bondi1952}, where mass infall rate ($\dot{M}_{\rm BH}$) is given as:
\begin{equation}
    \dot{M}_{\rm BH} = \frac{4\pi G^2m_{\rm H}n_{\rm H}}{(c_s^2 + v_{\rm rel.}^2)^{3/2}}M_{*}^{2}
\end{equation}
where $n_{\rm H}$, $c_s$, and $v_{\rm rel.}$ represent ambient gas number density, speed of sound, and the typical relative velocity between star and gas, respectively. 

Following this classical formulation and assuming highly supersonic turbulence in clouds, \cite{Padoan2025} derived the angular momentum expected for Bondi-Hoyle style infalling material within molecular clouds. Using arguments based on density versus size and velocity versus size relations observed in molecular clouds \citep{Larson1981,Solomon1987}, they derived time ($t$) dependence of $n_{\rm H}$ and $v_{\rm rel.}$ to be:
\begin{equation} \label{equ:params}
    n_{\rm H} = 9.6\times10^{3}\left(\frac{t}{\mathrm{Myr}}\right)^{-2}\,\mathrm{cm}^{-3};~ 
    v_{\rm rel.} = 0.75\left(\frac{t}{\mathrm{Myr}}\right)\,\mathrm{km}\,\mathrm{s}^{-1}
\end{equation}
Using these, they showed that the mass infall rate ($\dot{M}_{\rm BH}$) and specific angular momentum ($j_{\rm BH}$) at time $t$ of accreted material can be represented as:
\begin{align}
\dot{M}_{\rm BH} ={}& 1.3\times10^{-7}
\left(\frac{t}{\mathrm{Myr}}\right)^{-5}
\left(\frac{M_{*}}{M_{\odot}}\right)^{2}\,M_{\odot}\,\mathrm{yr^{-1}} \\
j_{\rm BH} ={}& 9.6\times10^{20}
\left(\frac{t}{\mathrm{Myr}}\right)^{-4}
\left(\frac{M_{*}}{M_{\odot}}\right)^{2}
\,\mathrm{cm^{2}\,s^{-1}}
\end{align}
These equations can be multiplied to estimate the total angular momentum infall rate. 

These analytical predictions agree well with the typical mass infall rates and specific angular momentum of infalling material found in numerical simulations of star-forming clouds with supersonic turbulence, as reported in \cite{Pelkonen2025}. Nevertheless, due to the highly structured nature of molecular clouds in the simulations, both quantities vary considerably among Class II sources of a given stellar mass: mass infall rates span three orders of magnitude, while specific angular momentum spans one order of magnitude. This is mainly because late infall is expected to be an episodic phenomenon, strongly depending on whether the young star is passing through an overdensity of gas or not.

The key physical parameters for the above derivations ($n_{\rm H}$, $v_{\rm rel.}$) are not observationally well constrained for Class II sources analyzed here. Regarding the gas densities, \citet{Rygl2013} found typical H$_2$ column densities of $\sim10^{22}$~cm$^{-2}$ around Class II sources in Lupus. Assuming the width of these clouds along the line-of-sight is $\sim1$~pc, the gas number densities will be $\sim3000$~cm$^{-3}$, which qualitatively agrees with the Equation \ref{equ:params}. Similarly, the typical velocity dispersion among Class II sources in nearby star-forming regions is of the order of $\sim1$~km~s$^{-1}$ \citep[e.g.,][]{Galli2020,Gupta2022}, which also qualitatively agrees with the relative velocities in Equation \ref{equ:params}. However, further studies are required to better constrain the gas densities and relative velocities in 3D for Class II sources at different ages, to ensure a fair comparison with these theoretical models of infalling material.

Class II phase of young stellar objects (YSOs) is expected to start after $\sim1$~Myr of the collapse of prestellar core, with their typical lifetimes of a few Myr \citep[e.g.,][]{Dunham2015}. These timescales are similar to the age of regions where most sources in our sample lie \citep[$\sim1$--3~Myr;][]{Ribas2015,Testi2022}, with Upper Sco sample from AGEPRO being a few Myr older. As the total mass and angular momentum of infalling material are time-integrated quantities, for a fair comparison we integrate them from 1--3~Myr, as shown in Figure \ref{fig:InfallComparison}. However, as the derived properties of infalling material strongly depends on the age of cluster, their age dependence is further discussed in Appendix \ref{app:age}.

\subsection{Observed streamers} \label{sec:infall_streamers}

\begin{table*}[ht]
\caption{Streamer properties}
\renewcommand{\arraystretch}{1.1}
\label{tab:streamers}
\centering
\begin{tabular}{lcccccl}
\hline\hline
Streamer &
$M_{\rm streamer}$ [$M_\odot$] &
$j_{\rm streamer}$ [cm$^2$ s$^{-1}$] &
$L_{\rm streamer}$ [g cm$^2$ s$^{-1}$] &
$M_*$ [$M_\odot$] &
Class &
Reference \\
\hline
S CrA & $2.10\times10^{-4}$ & $1.18\times10^{21}$ & $4.94\times10^{50}$ & 2.00 & II & \citet{Gupta2024} \\
HL Tau & $1.20\times10^{-5}$ & $9.07\times10^{20}$ & $2.16\times10^{49}$ & 2.10 & I/II & \citet{Gupta2024} \\
BHB1 & $1.60\times10^{-3}$ & $1.08\times10^{21}$ & $3.45\times10^{51}$ & 2.23 & I/II\tablefootmark{a} & \citet{Gupta2026} \\
L1489 & $1.83\times10^{-2}$ & $1.70\times10^{21}$ & $6.19\times10^{52}$ & 1.70 & I & \citet{Tanious2024} \\
FU Ori & $5.60\times10^{-7}$ & $1.51\times10^{21}$ & $1.68\times10^{48}$ & 1.50 & I & \citet{Hales2024} \\
M512 & $7.36\times10^{-2}$ & $2.54\times10^{21}$ & $3.72\times10^{53}$ & 0.15 & I & \citet{Cacciapuoti2024} \\
AB Aur S1 & -- & $4.04\times10^{21}$ & -- & 2.23 & II & \citet{Speedie2025} \\
AB Aur S2 & -- & $5.39\times10^{21}$ & -- & 2.23 & II & \citet{Speedie2025} \\
AB Aur S3 & -- & $1.72\times10^{21}$ & -- & 2.23 & II & \citet{Speedie2025} \\
DG Tau & -- & $9.06\times10^{19}$ & -- & 0.30 & I/II & \citet{Garufi2022} \\
IRAS 04302 S1 & -- & $6.45\times10^{20}$ & -- & 2.00 & I & \citet{Garufi2022} \\
Oph IRS 63 & $4.00\times10^{-2}$ & $1.87\times10^{21}$ & $1.49\times10^{53}$ & 0.50 & I & \citet{Flores2023} \\
Per-emb-50 & $1.20\times10^{-2}$ & $1.12\times10^{21}$ & $2.68\times10^{52}$ & 1.70 & I & \citet{Valdivia-Mena2022} \\
B5-IRS1 & -- & $4.95\times10^{20}$ & -- & 0.20 & I & \citet{Valdivia-Mena2023} \\
\hline
\end{tabular}
\tablefoot{
$M_{\rm streamer}$, $j_{\rm streamer}$, and $L_{\rm streamer}$ denote mass, specific angular momentum, and angular momentum of the streamer, respectively. As discussed in Section \ref{sec:infall}, mass and consequently, angular momentum of streamers are typically underestimated and should be treated as lower limits.
$M_*$ is the stellar mass of the central source. The table is also available in electronic form at the CDS via https://cdsarc.cds.unistra.fr/cgi-bin/qcat?J/A+A/VVV/AXX.\\
\tablefoottext{a}\,BHB1 may be a Class II source, appearing to be Class I/II due to the ongoing infall and an inclined disk \citep{Gupta2026}.
}
\end{table*}

For comparison with observations, we compile properties of streamers around Class I, Class I/II (flat spectrum), and Class II YSOs. 
For this comparison, we exclude streamers around Class 0 sources as they are more likely to be dynamically related to the prestellar cores and thus, may not contribute to misalignments \citep[e.g.,][]{Pelkonen2025}. 
Methods like Trajectory of Infalling Particles in Streamers around Young stars \citep[TIPSY;][]{Gupta2024} and Pineda Implementation of the Mendoza Streamline Model \citep[PIMS;][]{Pineda2020,Speedie2025} allows characterization of streamer dynamics by simultaneously fitting morphology and velocity gradients of streamers with theoretical trajectories of infalling material \citep[e.g.,][]{Mendoza2009}. 
The key assumption within these trajectories is that the only force acting on the infalling gas is the gravity from a point source (protostar). 
These methods can allow us to infer 3D orientations of streamers relative to the disks, as has been done for 
Class I source L1489 \citep{Tanious2024}, flat-spectrum (potentially a rejuvenated Class II) source BHB1 \citep{Gupta2026} and Class II source AB Aur \citep{Speedie2025}. The streamer around BHB1 shows a strong misalignment of $\sim69^{\circ}$, whereas the streamers around L1489 and AB Aur also exhibit moderate ($>10^{\circ}$) misalignments. In both of the methods, the best fit solutions allows us to infer the initial distance ($r_0$) and initial rotational velocity ($v_{\rm rot,0}$) of the infalling gas. Here, we used these quantities to estimate the specific angular momentum ($r_0\times v_{\rm rot,0}$) of the streamers, as reported in Table \ref{tab:streamers}. 

We note that derived orientations and specific angular momentum of streamers can be quite uncertain, as they rely on well constraining the 3D trajectory of infalling material. This requires systematically exploring all possible 3D orientations to identify a representative subset that can explain the observed streamer emission.
Moreover, depending on the observed length and orientation of the streamer, a systematic parameter search may still not sufficiently constrain the infalling parameters \citep[e.g., HL Tau in][]{Gupta2024}. BHB1 represents one of the best characterized streamers, where ALMA 12m array and ACA combined C$^{18}$O (2--1) observations of a clear streamer were analyzed using TIPSY \citep[][]{Gupta2026}. Even for this streamer, the derived misalignment ($69^{\circ}\pm34^{\circ}$) has significant uncertainties. The estimated specific angular momentum was $725\pm438$~au~km~s$^{-1}$, corresponding to an uncertainty of $\sim50\%$. This suggests that the derived values are likely good order-of-magnitude estimates, but the actual values may differ by a factor of a few. As uncertainties are rarely reported for streamer properties, we do not include them in the comparison presented in Section \ref{sec:results}.

In theory, the specific angular momentum of a streamer can be multiplied by its mass to estimate the total angular momentum \citep[e.g.,][]{Gupta2026}. However, in practice the mass estimation is quite complicated, as interferometric observations filter out large-scale emission. Furthermore, the limited field of view generally does not allow a comprehensive characterization of the total mass of the gas reservoirs feeding the streamers. Mass estimates are also typically derived under the assumption of optically thin emission, which may not always hold. Therefore, the reported streamer masses and corresponding angular momenta (see Table \ref{tab:streamers}) should be treated as lower limits. 

Roughly half of the streamers in Table \ref{tab:streamers} are missing mass estimates, as we could not find appropriate values in the corresponding manuscripts. We note that for the streamers around AB Aur, although the detailed dynamical analysis was performed by \cite{Speedie2025}, infall was already anticipated from molecular-line observations analyzed in \citet{Tang2012}, where the masses of streamer candidates were constrained to be $\sim10^{-7}$--$10^{-5}$~M$_{\odot}$, on the lower end compared to the rest of the sample presented here.

\section{Comparison of angular momentum} \label{sec:results}


\begin{figure}[htbp]
    \begin{subfigure}{0.5\textwidth}
        \centering
        \includegraphics[width=0.99\linewidth]{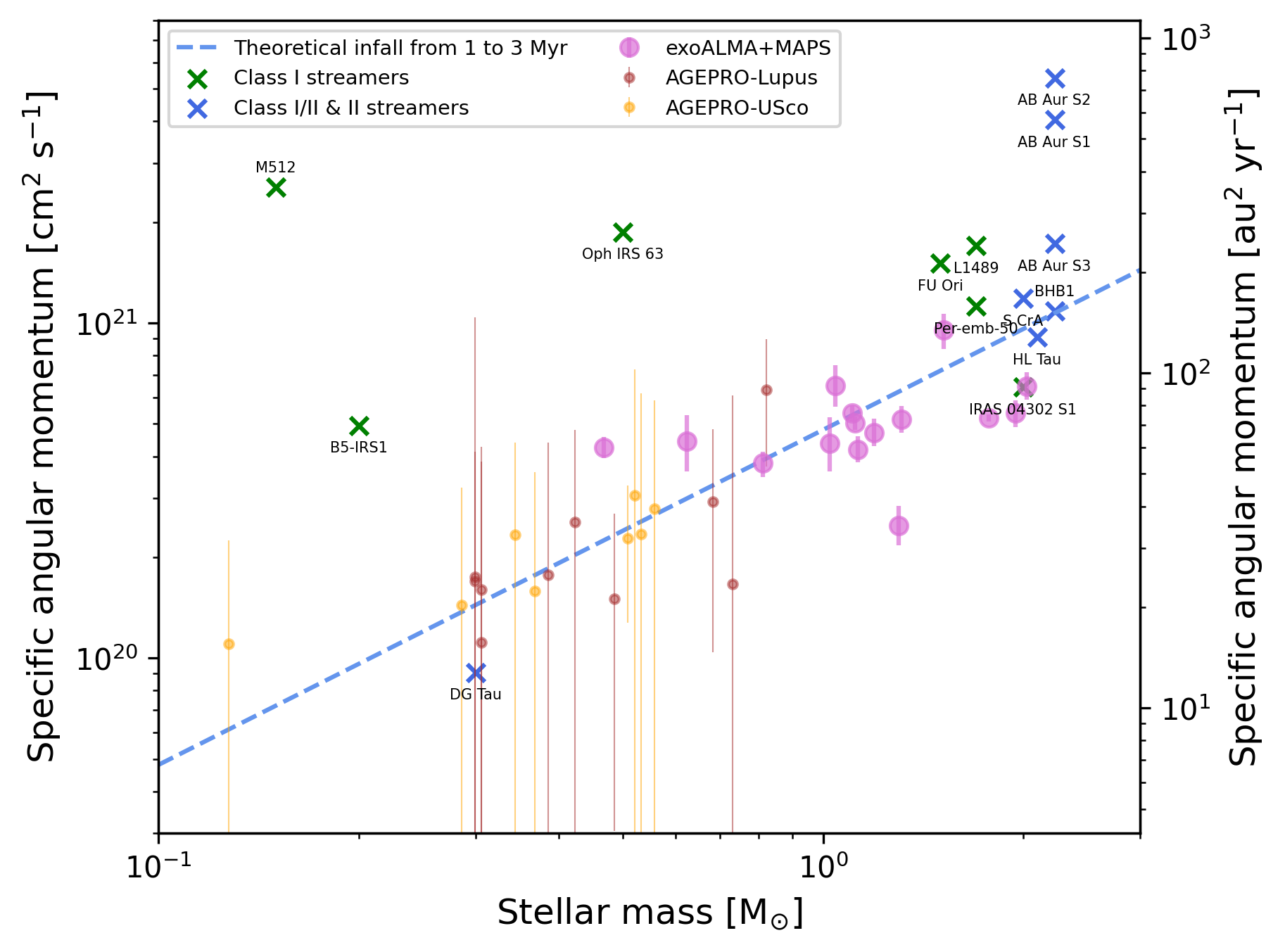}
        \label{fig:SpecificAngMomComparison}
    \end{subfigure}

    \begin{subfigure}{0.5\textwidth}
        \centering
        \includegraphics[width=0.99\linewidth]{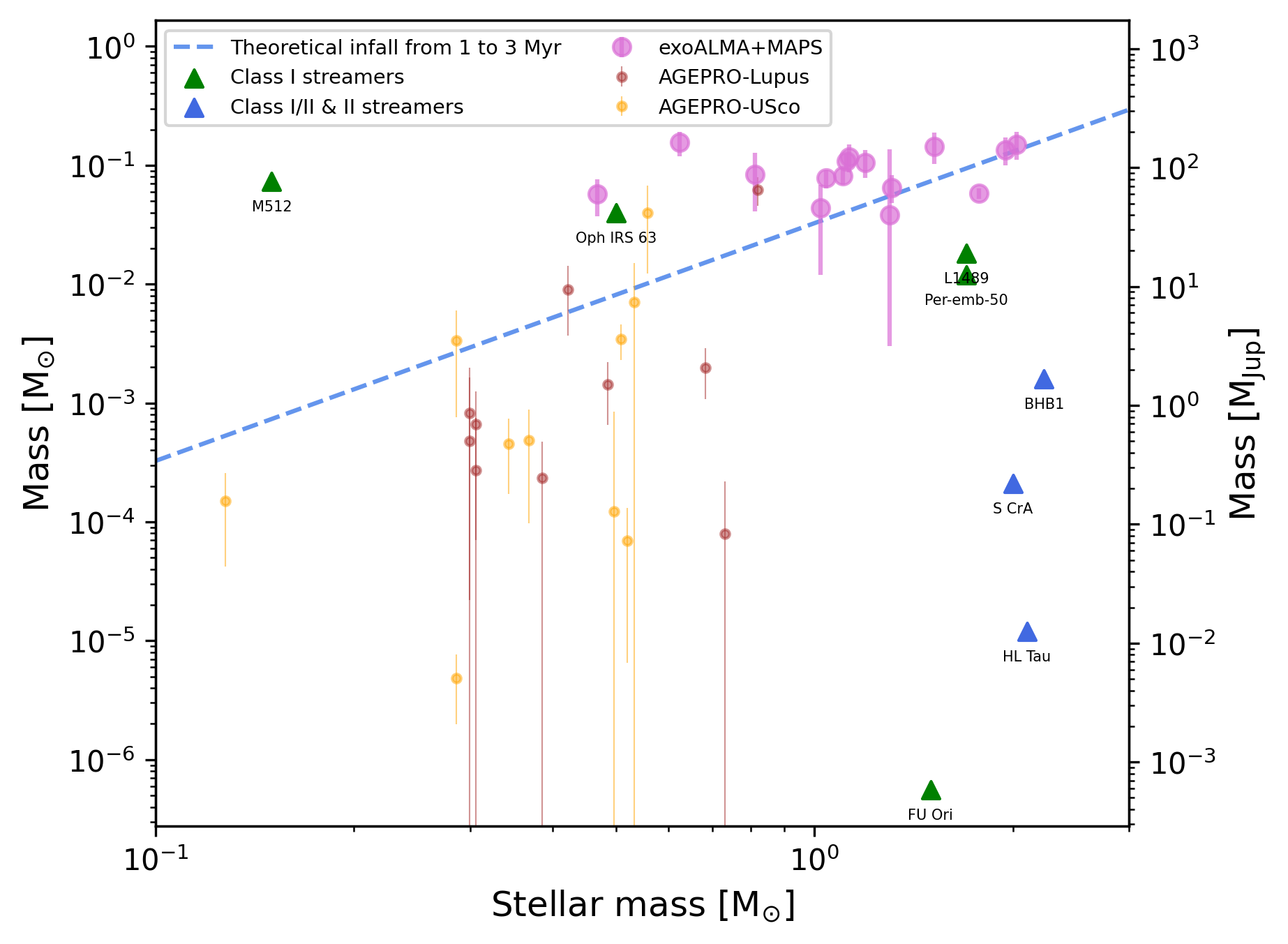}
        \label{fig:MassComparison}
    \end{subfigure}

    \begin{subfigure}{0.5\textwidth}
        \centering
        \includegraphics[width=0.99\linewidth]{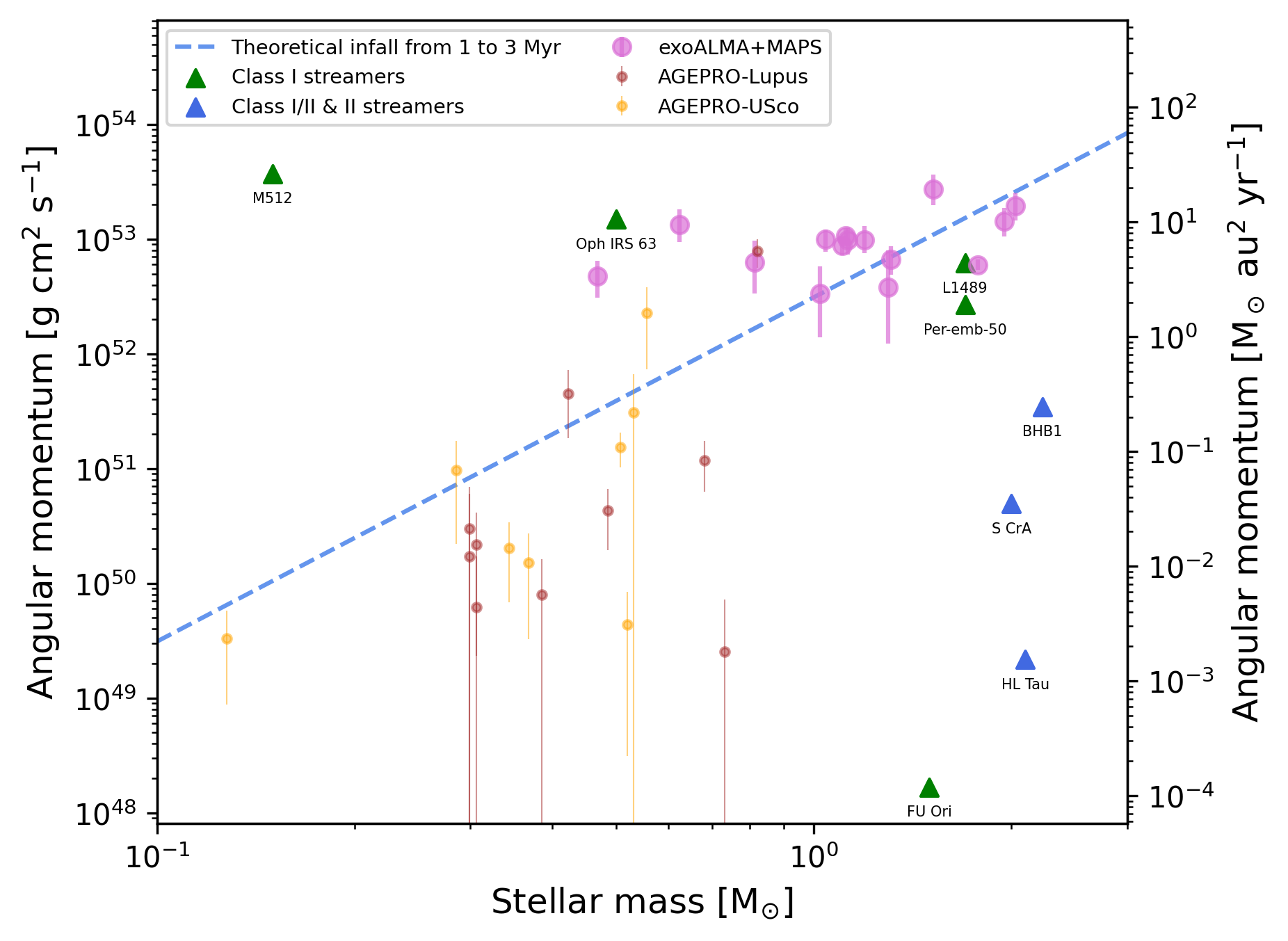}
        \label{fig:AngMomComparison}
    \end{subfigure}

    \caption{Specific angular momentum (top panel), mass (middle panel), and total angular momentum (lower panel) of disks and infalling material as a function of stellar mass. The purple circles represent disks with surface density profiles estimated in \citet{Martire2024} and \citet{Longarini2025}. The yellow (Upper Sco) and brown (Lupus) points represent disks observed within AGEPRO large program. The dashed line shows the theoretical infall relation from \citet{Padoan2025}. Observed streamers are represented as crosses in top panel and triangles (lower limits) in the middle and bottom panels, with green color representing Class I sources and blue color representing more evolved flat-spectrum or Class II sources.}
    \label{fig:InfallComparison}
\end{figure}

Figure \ref{fig:InfallComparison} compares the specific angular momentum (top panel), mass (middle panel), and total angular momentum (bottom panel) of disks (Section \ref{sec:disks}), observed streamers (Section \ref{sec:infall_streamers}), and theoretical predictions for late infall of material (Section \ref{sec:infall_pred}). As shown in the top panel, the specific angular momentum of disks is quite consistent with the theoretical expectations for infalling material. This agrees with the findings of \citet{Padoan2025}, where they estimated specific angular momentum of disks using dust radii measurements, assuming power-law surface density profiles \citep[see Eq. 29 in][]{Padoan2025}. These results further suggest that infalling material plays a role in regulating disk sizes which is more directly associated with specific angular momentum.
Moreover, the specific angular momentum of most streamers is comparable to, or slightly higher than, that of the disks. If the mass of these streamers is similar to, or slightly lower than, that of the disks, they could deliver angular momentum comparable to the disks and thus significantly misalign them \citep[e.g.,][]{Kuffmeier2021}.

As shown in the mass comparison (middle panel), for the streamers we generally only have a lower limit on the total mass, as discussed in Section \ref{sec:infall}. Although these lower limits for Class I streamers are comparable to disk masses, streamers around Class I/II and II sources are less massive by a few orders of magnitude. This large difference is unlikely to be accounted by the uncertainties on converting observed molecular-line emission to gas mass. 
However, if the bulk of the gas reservoirs feeding the streamers lies outside the field of view of the observations (typically a few thousand au), as has been suggested by some observational studies \citep[e.g.,][]{Valdivia-Mena2023}, then we may be underestimating mass of material that these streamers bring by such a huge margin. 
It is nonetheless interesting to note that even the most massive nearby disks have masses comparable to those theoretically expected to fall onto these systems. 
More typical disks lie below these theoretical expectations, suggesting that they could be readily replenished by late-stage infall. This discrepancy in masses could also be interpreted as this population of disks not being affected by infall. However, this may simply be due to infall being episodic with a large scatter \citep[e.g.,][]{Pelkonen2025} rather than continuous: some disks may have not yet undergone a significant infall episode, while others may have already experienced infall but subsequently lost the accreted mass to accretion onto the star or to planet formation, on timescales shorter than the disk's age.

The bottom panel shows the total angular momentum comparison. The predicted angular momentum of infalling material is comparable to that of the most massive disks and significantly higher than that of more typical AGEPRO disks. Although only lower limits are available for streamers, four of them (M512, Oph IRS 63, L1489, and Per-emb-50) already have angular momentum comparable to the disks, suggesting that they can effectively misalign these systems. However, all four are Class I streamers, and streamers at later stages may carry less angular momentum. 
It is plausible that a good fraction of disks are already misaligned with respect to the stellar spin by the streamers in the Class I phase.
However, this requires the early streamers themselves to be misaligned with respect to their disks, which is less likely in dynamically coherent pre-stellar cores. Numerical simulations presented in \citet{Pelkonen2025} suggest that the median misalignment between infalling material and the disk is $14^\circ$ at the Class I stage, significantly lower than $47^\circ$ at the Class II stage. It would still be interesting to include Class I disks in this comparison; however, their surface density profiles are not well constrained and they may not follow the same functional form as Class II disks.

As noted in Section \ref{sec:infall_pred}, the theoretical predictions on the properties of infalling material here are just to illustrate typical trends. As the cloud disk interactions are expected to be episodic \citep[e.g.,][]{Kuffmeier2024, Pelkonen2025}, there will a big diversity in the infall properties among the different sources. 

Overall, our results indicate that infalling material generally has specific angular momentum comparable to or higher than that of most disks. If streamers, especially those observed at later stages, are fed by gas reservoirs with masses comparable to or a bit lower than the disks, they will have total angular momentum similar to the disks and therefore, induce significant misalignments.

\section{Summary} \label{sec:summary} 

We present the first observationally anchored angular momentum values for planet-forming disks. We used surface density profiles inferred from dynamical modeling of high-resolution ALMA observations \citep[][]{Martire2024,Longarini2025} to directly measure the total angular momentum of 15 disks. 
In order to expand the analysis to more representative disks, we derived a physically motivated relationship to estimate angular momentum of disks based on their mass, the radius enclosing 90\% of the $^{13}$CO flux, and stellar mass. We applied this relation to disks observed in the AGEPRO Large Program, to estimate angular momentum of 18 more disks. 

Numerical simulations \citep[e.g.,][]{Kuffmeier2021} suggest that these disks can be misaligned compared to the stellar spin due to late-stage infall of material, if the infalling material has comparable angular momentum.
We inferred theoretically expected angular momentum of infalling material, approximated as a Bondi–Hoyle accretion \citep[e.g.,][]{Bondi1944,Bondi1952}, from analytical equations presented in \citet{Padoan2025}. The comparison of theoretical predictions and disk angular momentum in Figure \ref{fig:InfallComparison} shows the infall of material is expected to bring angular momentum comparable or exceeding most disks. This suggests that infall of material can influence disk sizes and orientation.

We also compiled observational estimates of initial positions, rotation velocities, and masses of streamers from the literature, and used them to constrain the specific and total angular momentum of infalling streamers. The specific angular momentum of streamers is in qualitative agreement with theoretical predictions and is higher than most disk values, suggesting that if they are fed by gas reservoirs of comparable mass, they can play an important role in inducing disk misalignment. However, as the mass estimates for streamers are lower limits due to the limited recoverable scales and field of view of interferometric observations, we can only derive lower limits on their angular momentum. Therefore, in order to observationally study the role of streamers in misaligning disks, we need to characterize the total mass of the gas reservoirs feeding them. Furthermore, the comparison presented here is limited by the small sample of streamers analyzed so far. Analysis of more streamers, in particular their angular momentum relative to their disks \citep[e.g.,][]{Gupta2026}, will allow us to further observationally test the scenario of infall-induced misalignment.

Overall, our results suggest that disks can be frequently misaligned due to interactions with the surrounding cloud, resulting in late-stage infall of material. This can naturally explain the misalignments observed in a significant fraction of disks and planetary systems.

\begin{acknowledgements}

We would like to thank Charles Law, Jonathan Williams, and Miguel Vioque for useful discussions regarding this idea of estimating angular momentum of disks. We are grateful to Dingshan Deng and Carolina Agurto-Gangas for their guidance in accessing AGEPRO data.

LIC and AG acknowledge support from NSF AST-2407547 and the David and Lucile Packard Foundation and the Virginia Institute of Theoretical Astronomy (VITA).
GR acknowledges support from the European Union (ERC Starting Grant DiscEvol, project number 101039651) and from Fondazione Cariplo, grant No. 2022-1217. CL and CJC have been supported by the UK Science and Technology Research Council (STFC) via the consolidated grant ST/W000997/1. Views and opinions expressed are, however, those of the author(s) only and do not necessarily reflect those of the European Union or the European Research Council. Neither the European Union nor the granting authority can be held responsible for them.
The work by M. K. is funded by the Independent Research Fund Denmark (DFF Sapere Aude Grant: 5251-00016B).

\end{acknowledgements}

\bibliographystyle{aa} 
\bibliography{refs} 

@ARTICLE{Gaudel2020,
       author = {{Gaudel}, M. and {Maury}, A.~J. and {Belloche}, A. and {Maret}, S. and {Andr{\'e}}, Ph. and {Hennebelle}, P. and {Galametz}, M. and {Testi}, L. and {Cabrit}, S. and {Palmeirim}, P. and {Ladjelate}, B. and {Codella}, C. and {Podio}, L.},
        title = "{Angular momentum profiles of Class 0 protostellar envelopes}",
      journal = {\aap},
         year = 2020,
        month = may,
       volume = {637},
          eid = {A92},
        pages = {A92},
          doi = {10.1051/0004-6361/201936364},
archivePrefix = {arXiv},
       eprint = {2001.10004},
 primaryClass = {astro-ph.SR},
       adsurl = {https://ui.adsabs.harvard.edu/abs/2020A&A...637A..92G}
}

@ARTICLE{Pineda2019,
       author = {{Pineda}, Jaime E. and {Zhao}, Bo and {Schmiedeke}, Anika and {Segura-Cox}, Dominique M. and {Caselli}, Paola and {Myers}, Philip C. and {Tobin}, John J. and {Dunham}, Michael},
        title = "{The Specific Angular Momentum Radial Profile in Dense Cores: Improved Initial Conditions for Disk Formation}",
      journal = {\apj},
         year = 2019,
        month = sep,
       volume = {882},
       number = {2},
          eid = {103},
        pages = {103},
          doi = {10.3847/1538-4357/ab2cd1},
archivePrefix = {arXiv},
       eprint = {1906.05578},
 primaryClass = {astro-ph.GA},
       adsurl = {https://ui.adsabs.harvard.edu/abs/2019ApJ...882..103P}
}

@ARTICLE{Yen2017,
       author = {{Yen}, Hsi-Wei and {Koch}, Patrick M. and {Takakuwa}, Shigehisa and {Krasnopolsky}, Ruben and {Ohashi}, Nagayoshi and {Aso}, Yusuke},
        title = "{Signs of Early-stage Disk Growth Revealed with ALMA}",
      journal = {\apj},
         year = 2017,
        month = jan,
       volume = {834},
       number = {2},
          eid = {178},
        pages = {178},
          doi = {10.3847/1538-4357/834/2/178},
archivePrefix = {arXiv},
       eprint = {1611.08416},
 primaryClass = {astro-ph.SR},
       adsurl = {https://ui.adsabs.harvard.edu/abs/2017ApJ...834..178Y}
}

@ARTICLE{Lee2016,
       author = {{Lee}, Chin-Fei and {Hwang}, Hsiang-Chih and {Li}, Zhi-Yun},
        title = "{Angular Momentum Loss in the Envelope-Disk Transition Region of the HH 111 Protostellar System: Evidence for Magnetic Braking?}",
      journal = {\apj},
         year = 2016,
        month = aug,
       volume = {826},
       number = {2},
          eid = {213},
        pages = {213},
          doi = {10.3847/0004-637X/826/2/213},
archivePrefix = {arXiv},
       eprint = {1605.08132},
 primaryClass = {astro-ph.GA},
       adsurl = {https://ui.adsabs.harvard.edu/abs/2016ApJ...826..213L}
}

@ARTICLE{Belloche2002,
       author = {{Belloche}, A. and {Andr{\'e}}, P. and {Despois}, D. and {Blinder}, S.},
        title = "{Molecular line study of the very young protostar IRAM 04191 in Taurus: infall, rotation, and outflow}",
      journal = {\aap},
         year = 2002,
        month = oct,
       volume = {393},
        pages = {927-947},
          doi = {10.1051/0004-6361:20021054},
archivePrefix = {arXiv},
       eprint = {astro-ph/0207287},
 primaryClass = {astro-ph},
       adsurl = {https://ui.adsabs.harvard.edu/abs/2002A&A...393..927B}
}

@ARTICLE{Ohashi1997,
       author = {{Ohashi}, Nagayoshi and {Hayashi}, Masahiko and {Ho}, Paul. T.~P. and {Momose}, Munetake and {Tamura}, Motohide and {Hirano}, Naomi and {Sargent}, Anneila I.},
        title = "{Rotation in the Protostellar Envelopes around IRAS 04169+2702 and IRAS 04365+2535: The Size Scale for Dynamical Collapse}",
      journal = {\apj},
         year = 1997,
        month = oct,
       volume = {488},
       number = {1},
        pages = {317-329},
          doi = {10.1086/304685},
       adsurl = {https://ui.adsabs.harvard.edu/abs/1997ApJ...488..317O}
}

@ARTICLE{Tang2012,
       author = {{Tang}, Y.-Wen and {Guilloteau}, S. and {Pi{\'e}tu}, V. and {Dutrey}, A. and {Ohashi}, N. and {Ho}, P.~T.~P.},
        title = "{The circumstellar disk of AB Aurigae: evidence for envelope accretion at late stages of star formation?}",
      journal = {\aap},
         year = 2012,
        month = nov,
       volume = {547},
          eid = {A84},
        pages = {A84},
          doi = {10.1051/0004-6361/201219414},
archivePrefix = {arXiv},
       eprint = {1209.1299},
 primaryClass = {astro-ph.GA},
       adsurl = {https://ui.adsabs.harvard.edu/abs/2012A&A...547A..84T}
}

@ARTICLE{Galli2020,
       author = {{Galli}, P.~A.~B. and {Bouy}, H. and {Olivares}, J. and {Miret-Roig}, N. and {Vieira}, R.~G. and {Sarro}, L.~M. and {Barrado}, D. and {Berihuete}, A. and {Bertout}, C. and {Bertin}, E. and {Cuillandre}, J.-C.},
        title = "{Lupus DANCe. Census of stars and 6D structure with Gaia-DR2 data}",
      journal = {\aap},
         year = 2020,
        month = nov,
       volume = {643},
          eid = {A148},
        pages = {A148},
          doi = {10.1051/0004-6361/202038717},
archivePrefix = {arXiv},
       eprint = {2010.00233},
 primaryClass = {astro-ph.SR},
       adsurl = {https://ui.adsabs.harvard.edu/abs/2020A&A...643A.148G}
}

@ARTICLE{Rygl2013,
       author = {{Rygl}, K.~L.~J. and {Benedettini}, M. and {Schisano}, E. and {Elia}, D. and {Molinari}, S. and {Pezzuto}, S. and {Andr{\'e}}, Ph. and {Bernard}, J.~P. and {White}, G.~J. and {Polychroni}, D. and {Bontemps}, S. and {Cox}, N.~L.~J. and {Di Francesco}, J. and {Facchini}, A. and {Fallscheer}, C. and {di Giorgio}, A.~M. and {Hennemann}, M. and {Hill}, T. and {K{\"o}nyves}, V. and {Minier}, V. and {Motte}, F. and {Nguyen-Luong}, Q. and {Peretto}, N. and {Pestalozzi}, M. and {Sadavoy}, S. and {Schneider}, N. and {Spinoglio}, L. and {Testi}, L. and {Ward-Thompson}, D.},
        title = "{Recent star formation in the Lupus clouds as seen by Herschel}",
      journal = {\aap},
         year = 2013,
        month = jan,
       volume = {549},
          eid = {L1},
        pages = {L1},
          doi = {10.1051/0004-6361/201219511},
archivePrefix = {arXiv},
       eprint = {1211.5232},
 primaryClass = {astro-ph.GA},
       adsurl = {https://ui.adsabs.harvard.edu/abs/2013A&A...549L...1R}
}

@ARTICLE{Larson1981,
       author = {{Larson}, R.~B.},
        title = "{Turbulence and star formation in molecular clouds.}",
      journal = {\mnras},
         year = 1981,
        month = mar,
       volume = {194},
        pages = {809-826},
          doi = {10.1093/mnras/194.4.809},
       adsurl = {https://ui.adsabs.harvard.edu/abs/1981MNRAS.194..809L}
}

@ARTICLE{Solomon1987,
       author = {{Solomon}, P.~M. and {Rivolo}, A.~R. and {Barrett}, J. and {Yahil}, A.},
        title = "{Mass, Luminosity, and Line Width Relations of Galactic Molecular Clouds}",
      journal = {\apj},
         year = 1987,
        month = aug,
       volume = {319},
        pages = {730},
          doi = {10.1086/165493},
       adsurl = {https://ui.adsabs.harvard.edu/abs/1987ApJ...319..730S}
}

@ARTICLE{Andrews2018,
       author = {{Andrews}, Sean M. and {Terrell}, Marie and {Tripathi}, Anjali and {Ansdell}, Megan and {Williams}, Jonathan P. and {Wilner}, David J.},
        title = "{Scaling Relations Associated with Millimeter Continuum Sizes in Protoplanetary Disks}",
      journal = {\apj},
         year = 2018,
        month = oct,
       volume = {865},
       number = {2},
          eid = {157},
        pages = {157},
          doi = {10.3847/1538-4357/aadd9f},
archivePrefix = {arXiv},
       eprint = {1808.10510},
 primaryClass = {astro-ph.EP},
       adsurl = {https://ui.adsabs.harvard.edu/abs/2018ApJ...865..157A}
}

@ARTICLE{Winter2025,
       author = {{Winter}, Andrew J. and {Benisty}, Myriam and {Izquierdo}, Andr{\'e}s F. and {Lodato}, Giuseppe and {Teague}, Richard and {Kimmig}, Carolin N. and {Andrews}, Sean M. and {Bae}, Jaehan and {Barraza-Alfaro}, Marcelo and {Cuello}, Nicol{\'a}s and {Curone}, Pietro and {Czekala}, Ian and {Facchini}, Stefano and {Fasano}, Daniele and {Hall}, Cassandra and {Hardiman}, Caitlyn and {Hilder}, Thomas and {Ilee}, John D. and {Fukagawa}, Misato and {Longarini}, Cristiano and {M{\'e}nard}, Fran{\c{c}}ois and {Orihara}, Ryuta and {Pinte}, Christophe and {Price}, Daniel J. and {Rosotti}, Giovanni and {Stadler}, Jochen and {Wilner}, David J. and {W{\"o}lfer}, Lisa and {Yen}, Hsi-Wei and {Yoshida}, Tomohiro C. and {Zawadzki}, Brianna},
        title = "{exoALMA. XVIII. Interpreting Large-scale Kinematic Structures as Moderate Warping}",
      journal = {\apjl},
         year = 2025,
        month = sep,
       volume = {990},
       number = {1},
          eid = {L10},
        pages = {L10},
          doi = {10.3847/2041-8213/adf113},
archivePrefix = {arXiv},
       eprint = {2507.11669},
 primaryClass = {astro-ph.EP},
       adsurl = {https://ui.adsabs.harvard.edu/abs/2025ApJ...990L..10W}
}

@ARTICLE{Kimmig2025,
       author = {{Kimmig}, Carolin N. and {Villenave}, Marion},
        title = "{Asymmetric signatures of warps in edge-on disks}",
      journal = {\aap},
         year = 2025,
        month = jun,
       volume = {698},
          eid = {A146},
        pages = {A146},
          doi = {10.1051/0004-6361/202453313},
archivePrefix = {arXiv},
       eprint = {2504.05399},
 primaryClass = {astro-ph.SR},
       adsurl = {https://ui.adsabs.harvard.edu/abs/2025A&A...698A.146K}
}

@ARTICLE{Dunham2015,
       author = {{Dunham}, Michael M. and {Allen}, Lori E. and {Evans}, II, Neal J. and {Broekhoven-Fiene}, Hannah and {Cieza}, Lucas A. and {Di Francesco}, James and {Gutermuth}, Robert A. and {Harvey}, Paul M. and {Hatchell}, Jennifer and {Heiderman}, Amanda and {Huard}, Tracy L. and {Johnstone}, Doug and {Kirk}, Jason M. and {Matthews}, Brenda C. and {Miller}, Jennifer F. and {Peterson}, Dawn E. and {Young}, Kaisa E.},
        title = "{Young Stellar Objects in the Gould Belt}",
      journal = {\apjs},
         year = 2015,
        month = sep,
       volume = {220},
       number = {1},
          eid = {11},
        pages = {11},
          doi = {10.1088/0067-0049/220/1/11},
archivePrefix = {arXiv},
       eprint = {1508.03199},
 primaryClass = {astro-ph.GA},
       adsurl = {https://ui.adsabs.harvard.edu/abs/2015ApJS..220...11D}
}

@ARTICLE{Testi2022,
       author = {{Testi}, L. and {Natta}, A. and {Manara}, C.~F. and {de Gregorio Monsalvo}, I. and {Lodato}, G. and {Lopez}, C. and {Muzic}, K. and {Pascucci}, I. and {Sanchis}, E. and {Miranda}, A. Santamaria and {Scholz}, A. and {De Simone}, M. and {Williams}, J.~P.},
        title = "{The protoplanetary disk population in the {\ensuremath{\rho}}-Ophiuchi region L1688 and the time evolution of Class II YSOs}",
      journal = {\aap},
         year = 2022,
        month = jul,
       volume = {663},
          eid = {A98},
        pages = {A98},
          doi = {10.1051/0004-6361/202141380},
archivePrefix = {arXiv},
       eprint = {2201.04079},
 primaryClass = {astro-ph.SR},
       adsurl = {https://ui.adsabs.harvard.edu/abs/2022A&A...663A..98T}
}

@ARTICLE{Ribas2015,
       author = {{Ribas}, {\'A}lvaro and {Bouy}, Herv{\'e} and {Mer{\'\i}n}, Bruno},
        title = "{Protoplanetary disk lifetimes vs. stellar mass and possible implications for giant planet populations}",
      journal = {\aap},
         year = 2015,
        month = apr,
       volume = {576},
          eid = {A52},
        pages = {A52},
          doi = {10.1051/0004-6361/201424846},
archivePrefix = {arXiv},
       eprint = {1502.00631},
 primaryClass = {astro-ph.SR},
       adsurl = {https://ui.adsabs.harvard.edu/abs/2015A&A...576A..52R}
}

@ARTICLE{Gaia2016,
       author = {{Gaia Collaboration} and {Prusti}, T. and {de Bruijne}, J.~H.~J. and {Brown}, A.~G.~A. and {Vallenari}, A. and {Babusiaux}, C. and {Bailer-Jones}, C.~A.~L. and {Bastian}, U. and {Biermann}, M. and {Evans}, D.~W. and {Eyer}, L. and {Jansen}, F. and {Jordi}, C. and {Klioner}, S.~A. and {Lammers}, U. and {Lindegren}, L. and {Luri}, X. and {Mignard}, F. and {Milligan}, D.~J. and {Panem}, C. and {Poinsignon}, V. and {Pourbaix}, D. and {Randich}, S. and {Sarri}, G. and {Sartoretti}, P. and {Siddiqui}, H.~I. and {Soubiran}, C. and {Valette}, V. and {van Leeuwen}, F. and {Walton}, N.~A. and {Aerts}, C. and {Arenou}, F. and {Cropper}, M. and {Drimmel}, R. and {H{\o}g}, E. and {Katz}, D. and {Lattanzi}, M.~G. and {O'Mullane}, W. and {Grebel}, E.~K. and {Holland}, A.~D. and {Huc}, C. and {Passot}, X. and {Bramante}, L. and {Cacciari}, C. and {Casta{\~n}eda}, J. and {Chaoul}, L. and {Cheek}, N. and {De Angeli}, F. and {Fabricius}, C. and {Guerra}, R. and {Hern{\'a}ndez}, J. and {Jean-Antoine-Piccolo}, A. and {Masana}, E. and {Messineo}, R. and {Mowlavi}, N. and {Nienartowicz}, K. and {Ord{\'o}{\~n}ez-Blanco}, D. and {Panuzzo}, P. and {Portell}, J. and {Richards}, P.~J. and {Riello}, M. and {Seabroke}, G.~M. and {Tanga}, P. and {Th{\'e}venin}, F. and {Torra}, J. and {Els}, S.~G. and {Gracia-Abril}, G. and {Comoretto}, G. and {Garcia-Reinaldos}, M. and {Lock}, T. and {Mercier}, E. and {Altmann}, M. and {Andrae}, R. and {Astraatmadja}, T.~L. and {Bellas-Velidis}, I. and {Benson}, K. and {Berthier}, J. and {Blomme}, R. and {Busso}, G. and {Carry}, B. and {Cellino}, A. and {Clementini}, G. and {Cowell}, S. and {Creevey}, O. and {Cuypers}, J. and {Davidson}, M. and {De Ridder}, J. and {de Torres}, A. and {Delchambre}, L. and {Dell'Oro}, A. and {Ducourant}, C. and {Fr{\'e}mat}, Y. and {Garc{\'\i}a-Torres}, M. and {Gosset}, E. and {Halbwachs}, J.-L. and {Hambly}, N.~C. and {Harrison}, D.~L. and {Hauser}, M. and {Hestroffer}, D. and {Hodgkin}, S.~T. and {Huckle}, H.~E. and {Hutton}, A. and {Jasniewicz}, G. and {Jordan}, S. and {Kontizas}, M. and {Korn}, A.~J. and {Lanzafame}, A.~C. and {Manteiga}, M. and {Moitinho}, A. and {Muinonen}, K. and {Osinde}, J. and {Pancino}, E. and {Pauwels}, T. and {Petit}, J.-M. and {Recio-Blanco}, A. and {Robin}, A.~C. and {Sarro}, L.~M. and {Siopis}, C. and {Smith}, M. and {Smith}, K.~W. and {Sozzetti}, A. and {Thuillot}, W. and {van Reeven}, W. and {Viala}, Y. and {Abbas}, U. and {Abreu Aramburu}, A. and {Accart}, S. and {Aguado}, J.~J. and {Allan}, P.~M. and {Allasia}, W. and {Altavilla}, G. and {{\'A}lvarez}, M.~A. and {Alves}, J. and {Anderson}, R.~I. and {Andrei}, A.~H. and {Anglada Varela}, E. and {Antiche}, E. and {Antoja}, T. and {Ant{\'o}n}, S. and {Arcay}, B. and {Atzei}, A. and {Ayache}, L. and {Bach}, N. and {Baker}, S.~G. and {Balaguer-N{\'u}{\~n}ez}, L. and {Barache}, C. and {Barata}, C. and {Barbier}, A. and {Barblan}, F. and {Baroni}, M. and {Barrado y Navascu{\'e}s}, D. and {Barros}, M. and {Barstow}, M.~A. and {Becciani}, U. and {Bellazzini}, M. and {Bellei}, G. and {Bello Garc{\'\i}a}, A. and {Belokurov}, V. and {Bendjoya}, P. and {Berihuete}, A. and {Bianchi}, L. and {Bienaym{\'e}}, O. and {Billebaud}, F. and {Blagorodnova}, N. and {Blanco-Cuaresma}, S. and {Boch}, T. and {Bombrun}, A. and {Borrachero}, R. and {Bouquillon}, S. and {Bourda}, G. and {Bouy}, H. and {Bragaglia}, A. and {Breddels}, M.~A. and {Brouillet}, N. and {Br{\"u}semeister}, T. and {Bucciarelli}, B. and {Budnik}, F. and {Burgess}, P. and {Burgon}, R. and {Burlacu}, A. and {Busonero}, D. and {Buzzi}, R. and {Caffau}, E. and {Cambras}, J. and {Campbell}, H. and {Cancelliere}, R. and {Cantat-Gaudin}, T. and {Carlucci}, T. and {Carrasco}, J.~M. and {Castellani}, M. and {Charlot}, P. and {Charnas}, J. and {Charvet}, P. and {Chassat}, F. and {Chiavassa}, A. and {Clotet}, M. and {Cocozza}, G. and {Collins}, R.~S. and {Collins}, P. and {Costigan}, G.},
        title = "{The Gaia mission}",
      journal = {\aap},
         year = 2016,
        month = nov,
       volume = {595},
          eid = {A1},
        pages = {A1},
          doi = {10.1051/0004-6361/201629272},
archivePrefix = {arXiv},
       eprint = {1609.04153},
 primaryClass = {astro-ph.IM},
       adsurl = {https://ui.adsabs.harvard.edu/abs/2016A&A...595A...1G}
}

@ARTICLE{Stefansson2025,
       author = {{Stef{\'a}nsson}, Gudmundur and {Mahadevan}, Suvrath and {Winn}, Joshua N. and {Marcussen}, Marcus L. and {Kanodia}, Shubham and {Albrecht}, Simon and {Fitzmaurice}, Evan and {Mikulskyt{\.{e}}}, On{\.{e}} and {Ca{\~n}as}, Caleb I. and {Espinoza-Retamal}, Juan I. and {Zwart}, Yiri and {Krolikowski}, Daniel M. and {Hotnisky}, Andrew and {Robertson}, Paul and {Alvarado-Montes}, Jaime A. and {Bender}, Chad F. and {Blake}, Cullen H. and {Callingham}, J.~R. and {Cochran}, William D. and {Delamer}, Megan and {Diddams}, Scott A. and {Dong}, Jiayin and {Fernandes}, Rachel B. and {Giovinazzi}, Mark R. and {Halverson}, Samuel and {Libby-Roberts}, Jessica and {Logsdon}, Sarah E. and {McElwain}, Michael W. and {Ninan}, Joe P. and {Rajagopal}, Jayadev and {Reji}, Varghese and {Roy}, Arpita and {Schwab}, Christian and {Wright}, Jason T.},
        title = "{Gaia-4b and 5b: Radial Velocity Confirmation of Gaia Astrometric Orbital Solutions Reveal a Massive Planet and a Brown Dwarf Orbiting Low-mass Stars}",
      journal = {\aj},
         year = 2025,
        month = feb,
       volume = {169},
       number = {2},
          eid = {107},
        pages = {107},
          doi = {10.3847/1538-3881/ada9e1},
archivePrefix = {arXiv},
       eprint = {2410.05654},
 primaryClass = {astro-ph.EP},
       adsurl = {https://ui.adsabs.harvard.edu/abs/2025AJ....169..107S}
}

@ARTICLE{Borucki2010,
       author = {{Borucki}, William J. and {Koch}, David and {Basri}, Gibor and {Batalha}, Natalie and {Brown}, Timothy and {Caldwell}, Douglas and {Caldwell}, John and {Christensen-Dalsgaard}, J{\o}rgen and {Cochran}, William D. and {DeVore}, Edna and {Dunham}, Edward W. and {Dupree}, Andrea K. and {Gautier}, Thomas N. and {Geary}, John C. and {Gilliland}, Ronald and {Gould}, Alan and {Howell}, Steve B. and {Jenkins}, Jon M. and {Kondo}, Yoji and {Latham}, David W. and {Marcy}, Geoffrey W. and {Meibom}, S{\o}ren and {Kjeldsen}, Hans and {Lissauer}, Jack J. and {Monet}, David G. and {Morrison}, David and {Sasselov}, Dimitar and {Tarter}, Jill and {Boss}, Alan and {Brownlee}, Don and {Owen}, Toby and {Buzasi}, Derek and {Charbonneau}, David and {Doyle}, Laurance and {Fortney}, Jonathan and {Ford}, Eric B. and {Holman}, Matthew J. and {Seager}, Sara and {Steffen}, Jason H. and {Welsh}, William F. and {Rowe}, Jason and {Anderson}, Howard and {Buchhave}, Lars and {Ciardi}, David and {Walkowicz}, Lucianne and {Sherry}, William and {Horch}, Elliott and {Isaacson}, Howard and {Everett}, Mark E. and {Fischer}, Debra and {Torres}, Guillermo and {Johnson}, John Asher and {Endl}, Michael and {MacQueen}, Phillip and {Bryson}, Stephen T. and {Dotson}, Jessie and {Haas}, Michael and {Kolodziejczak}, Jeffrey and {Van Cleve}, Jeffrey and {Chandrasekaran}, Hema and {Twicken}, Joseph D. and {Quintana}, Elisa V. and {Clarke}, Bruce D. and {Allen}, Christopher and {Li}, Jie and {Wu}, Haley and {Tenenbaum}, Peter and {Verner}, Ekaterina and {Bruhweiler}, Frederick and {Barnes}, Jason and {Prsa}, Andrej},
        title = "{Kepler Planet-Detection Mission: Introduction and First Results}",
      journal = {Science},
         year = 2010,
        month = feb,
       volume = {327},
       number = {5968},
        pages = {977},
          doi = {10.1126/science.1185402},
       adsurl = {https://ui.adsabs.harvard.edu/abs/2010Sci...327..977B}
}

@ARTICLE{Morbidelli2016,
       author = {{Morbidelli}, Alessandro and {Raymond}, Sean N.},
        title = "{Challenges in planet formation}",
      journal = {Journal of Geophysical Research (Planets)},
         year = 2016,
        month = oct,
       volume = {121},
       number = {10},
        pages = {1962-1980},
          doi = {10.1002/2016JE005088},
archivePrefix = {arXiv},
       eprint = {1610.07202},
 primaryClass = {astro-ph.EP},
       adsurl = {https://ui.adsabs.harvard.edu/abs/2016JGRE..121.1962M}
}

@ARTICLE{Tabone2022,
       author = {{Tabone}, Beno{\^\i}t and {Rosotti}, Giovanni P. and {Cridland}, Alexander J. and {Armitage}, Philip J. and {Lodato}, Giuseppe},
        title = "{Secular evolution of MHD wind-driven discs: analytical solutions in the expanded {\ensuremath{\alpha}}-framework}",
      journal = {\mnras},
         year = 2022,
        month = may,
       volume = {512},
       number = {2},
        pages = {2290-2309},
          doi = {10.1093/mnras/stab3442},
archivePrefix = {arXiv},
       eprint = {2111.10145},
 primaryClass = {astro-ph.SR},
       adsurl = {https://ui.adsabs.harvard.edu/abs/2022MNRAS.512.2290T}
}

@ARTICLE{Somigliana2023,
       author = {{Somigliana}, Alice and {Testi}, Leonardo and {Rosotti}, Giovanni and {Toci}, Claudia and {Lodato}, Giuseppe and {Tabone}, Beno{\^\i}t and {Manara}, Carlo F. and {Tazzari}, Marco},
        title = "{The Time Evolution of \{M\}\_\{d\}/\textbackslashdot\{M\} in Protoplanetary Disks as a Way to Disentangle between Viscosity and MHD Winds}",
      journal = {\apjl},
         year = 2023,
        month = sep,
       volume = {954},
       number = {1},
          eid = {L13},
        pages = {L13},
          doi = {10.3847/2041-8213/acf048},
archivePrefix = {arXiv},
       eprint = {2309.04496},
 primaryClass = {astro-ph.EP},
       adsurl = {https://ui.adsabs.harvard.edu/abs/2023ApJ...954L..13S}
}

@ARTICLE{Lesur2021,
       author = {{Lesur}, G.},
        title = "{Magnetohydrodynamics of protoplanetary discs}",
      journal = {Journal of Plasma Physics},
         year = 2021,
        month = feb,
       volume = {87},
       number = {1},
          eid = {205870101},
        pages = {205870101},
          doi = {10.1017/S0022377820001002},
archivePrefix = {arXiv},
       eprint = {2007.15967},
 primaryClass = {astro-ph.SR},
       adsurl = {https://ui.adsabs.harvard.edu/abs/2021JPlPh..87a2001P}
}

@ARTICLE{Rosotti2023,
       author = {{Rosotti}, Giovanni P.},
        title = "{Empirical constraints on turbulence in proto-planetary discs}",
      journal = {\nar},
         year = 2023,
        month = jun,
       volume = {96},
          eid = {101674},
        pages = {101674},
          doi = {10.1016/j.newar.2023.101674},
archivePrefix = {arXiv},
       eprint = {2302.01433},
 primaryClass = {astro-ph.EP},
       adsurl = {https://ui.adsabs.harvard.edu/abs/2023NewAR..9601674R}
}

@INPROCEEDINGS{Manara2023,
       author = {{Manara}, C.~F. and {Ansdell}, M. and {Rosotti}, G.~P. and {Hughes}, A.~M. and {Armitage}, P.~J. and {Lodato}, G. and {Williams}, J.~P.},
        title = "{Demographics of Young Stars and their Protoplanetary Disks: Lessons Learned on Disk Evolution and its Connection to Planet Formation}",
    booktitle = {Protostars and Planets VII},
         year = 2023,
       editor = {{Inutsuka}, S. and {Aikawa}, Y. and {Muto}, T. and {Tomida}, K. and {Tamura}, M.},
       series = {Astronomical Society of the Pacific Conference Series},
       volume = {534},
        month = jul,
        pages = {539},
          doi = {10.48550/arXiv.2203.09930},
archivePrefix = {arXiv},
       eprint = {2203.09930},
 primaryClass = {astro-ph.SR},
       adsurl = {https://ui.adsabs.harvard.edu/abs/2023ASPC..534..539M}
}

@ARTICLE{joo12,
      author = {{Joos}, M. and {Hennebelle}, P. and {Ciardi}, A.},
        title = "{Protostellar disk formation and transport of angular momentum during magnetized core collapse}",
      journal = {\aap},
         year = 2012,
        month = jul,
      volume = {543},
          eid = {A128},
        pages = {A128},
          doi = {10.1051/0004-6361/201118730},
archivePrefix = {arXiv},
      eprint = {1203.1193},
 primaryClass = {astro-ph.SR},
      adsurl = {https://ui.adsabs.harvard.edu/abs/2012A&A...543A.128J}
}

@ARTICLE{all03,
       author = {{Allen}, Anthony and {Li}, Zhi-Yun and {Shu}, Frank H.},
        title = "{Collapse of Magnetized Singular Isothermal Toroids. II. Rotation and Magnetic Braking}",
      journal = {\apj},
         year = 2003,
        month = dec,
       volume = {599},
       number = {1},
        pages = {363-379},
          doi = {10.1086/379243},
archivePrefix = {arXiv},
       eprint = {astro-ph/0311377},
 primaryClass = {astro-ph},
       adsurl = {https://ui.adsabs.harvard.edu/abs/2003ApJ...599..363A}
}

@ARTICLE{gal06,
       author = {{Galli}, Daniele and {Lizano}, Susana and {Shu}, Frank H. and {Allen}, Anthony},
        title = "{Gravitational Collapse of Magnetized Clouds. I. Ideal Magnetohydrodynamic Accretion Flow}",
      journal = {\apj},
         year = 2006,
        month = aug,
       volume = {647},
       number = {1},
        pages = {374-381},
          doi = {10.1086/505257},
archivePrefix = {arXiv},
       eprint = {astro-ph/0604573},
 primaryClass = {astro-ph},
       adsurl = {https://ui.adsabs.harvard.edu/abs/2006ApJ...647..374G}
}

@ARTICLE{mel08,
       author = {{Mellon}, Richard R. and {Li}, Zhi-Yun},
        title = "{Magnetic Braking and Protostellar Disk Formation: The Ideal MHD Limit}",
      journal = {\apj},
         year = 2008,
        month = jul,
       volume = {681},
       number = {2},
        pages = {1356-1376},
          doi = {10.1086/587542},
archivePrefix = {arXiv},
       eprint = {0709.0445},
 primaryClass = {astro-ph},
       adsurl = {https://ui.adsabs.harvard.edu/abs/2008ApJ...681.1356M}
}

@ARTICLE{Gupta2022_mag,
       author = {{Gupta}, Aashish and {Yen}, Hsi-Wei and {Koch}, Patrick and {Bastien}, Pierre and {Bourke}, Tyler L. and {Chung}, Eun Jung and {Hasegawa}, Tetsuo and {Hull}, Charles L.~H. and {Inutsuka}, Shu-ichiro and {Kwon}, Jungmi and {Kwon}, Woojin and {Lai}, Shih-Ping and {Lee}, Chang Won and {Lee}, Chin-Fei and {Pattle}, Kate and {Qiu}, Keping and {Tahani}, Mehrnoosh and {Tamura}, Motohide and {Ward-Thompson}, Derek},
        title = "{Effects of Magnetic Field Orientations in Dense Cores on Gas Kinematics in Protostellar Envelopes}",
      journal = {\apj},
         year = 2022,
        month = may,
       volume = {930},
       number = {1},
          eid = {67},
        pages = {67},
          doi = {10.3847/1538-4357/ac63bc},
archivePrefix = {arXiv},
       eprint = {2204.05636},
 primaryClass = {astro-ph.SR},
       adsurl = {https://ui.adsabs.harvard.edu/abs/2022ApJ...930...67G}
}

@ARTICLE{Jiang2022,
       author = {{Jiang}, Jonathan H. and {Burn}, Remo and {Ji}, Xuan and {Fahy}, Kristen A. and {Eggenberger}, Patrick},
        title = "{Angular Momentum Distributions for Observed and Modeled Exoplanetary Systems}",
      journal = {\apj},
         year = 2022,
        month = jan,
       volume = {924},
       number = {2},
          eid = {118},
        pages = {118},
          doi = {10.3847/1538-4357/ac3242},
archivePrefix = {arXiv},
       eprint = {2108.02890},
 primaryClass = {astro-ph.EP},
       adsurl = {https://ui.adsabs.harvard.edu/abs/2022ApJ...924..118J}
}

@ARTICLE{Tatematsu2016,
       author = {{Tatematsu}, Ken'ichi and {Ohashi}, Satoshi and {Sanhueza}, Patricio and {Nguyen Luong}, Quang and {Umemoto}, Tomofumi and {Mizuno}, Norikazu},
        title = "{Angular momentum of the N$_{2}$H$^{+}$ cores in the Orion A cloud}",
      journal = {\pasj},
         year = 2016,
        month = apr,
       volume = {68},
       number = {2},
          eid = {24},
        pages = {24},
          doi = {10.1093/pasj/psw002},
archivePrefix = {arXiv},
       eprint = {1601.00362},
 primaryClass = {astro-ph.GA},
       adsurl = {https://ui.adsabs.harvard.edu/abs/2016PASJ...68...24T}
}

@ARTICLE{Garufi2022,
       author = {{Garufi}, A. and {Podio}, L. and {Codella}, C. and {Segura-Cox}, D. and {Vander Donckt}, M. and {Mercimek}, S. and {Bacciotti}, F. and {Fedele}, D. and {Kasper}, M. and {Pineda}, J.~E. and {Humphreys}, E. and {Testi}, L.},
        title = "{ALMA chemical survey of disk-outflow sources in Taurus (ALMA-DOT). VI. Accretion shocks in the disk of DG Tau and HL Tau}",
      journal = {\aap},
         year = 2022,
        month = feb,
       volume = {658},
          eid = {A104},
        pages = {A104},
          doi = {10.1051/0004-6361/202141264},
archivePrefix = {arXiv},
       eprint = {2110.13820},
 primaryClass = {astro-ph.GA},
       adsurl = {https://ui.adsabs.harvard.edu/abs/2022A&A...658A.104G}
}

@ARTICLE{Valdivia-Mena2022,
       author = {{Valdivia-Mena}, M.~T. and {Pineda}, J.~E. and {Segura-Cox}, D.~M. and {Caselli}, P. and {Neri}, R. and {L{\'o}pez-Sepulcre}, A. and {Cunningham}, N. and {Bouscasse}, L. and {Semenov}, D. and {Henning}, Th. and {Pi{\'e}tu}, V. and {Chapillon}, E. and {Dutrey}, A. and {Fuente}, A. and {Guilloteau}, S. and {Hsieh}, T.~H. and {Jim{\'e}nez-Serra}, I. and {Marino}, S. and {Maureira}, M.~J. and {Smirnov-Pinchukov}, G.~V. and {Tafalla}, M. and {Zhao}, B.},
        title = "{PRODIGE - envelope to disk with NOEMA. I. A 3000 au streamer feeding a Class I protostar}",
      journal = {\aap},
         year = 2022,
        month = nov,
       volume = {667},
          eid = {A12},
        pages = {A12},
          doi = {10.1051/0004-6361/202243310},
archivePrefix = {arXiv},
       eprint = {2208.01023},
 primaryClass = {astro-ph.GA},
       adsurl = {https://ui.adsabs.harvard.edu/abs/2022A&A...667A..12V}
}

@ARTICLE{Valdivia-Mena2023,
       author = {{Valdivia-Mena}, M.~T. and {Pineda}, J.~E. and {Segura-Cox}, D.~M. and {Caselli}, P. and {Schmiedeke}, A. and {Choudhury}, S. and {Offner}, S.~S.~R. and {Neri}, R. and {Goodman}, A. and {Fuller}, G.~A.},
        title = "{Flow of gas detected from beyond the filaments to protostellar scales in Barnard 5★}",
      journal = {\aap},
         year = 2023,
        month = sep,
       volume = {677},
          eid = {A92},
        pages = {A92},
          doi = {10.1051/0004-6361/202346357},
archivePrefix = {arXiv},
       eprint = {2307.14337},
 primaryClass = {astro-ph.GA},
       adsurl = {https://ui.adsabs.harvard.edu/abs/2023A&A...677A..92V}
}

@ARTICLE{Cacciapuoti2024,
       author = {{Cacciapuoti}, L. and {Macias}, E. and {Gupta}, A. and {Testi}, L. and {Miotello}, A. and {Espaillat}, C. and {K{\"u}ffmeier}, M. and {van Terwisga}, S. and {Tobin}, J. and {Grant}, S. and {Manara}, C.~F. and {Segura-Cox}, D. and {Wendeborn}, J. and {Klessen}, R.~S. and {Maury}, A.~J. and {Lebreuilly}, U. and {Hennebelle}, P. and {Molinari}, S.},
        title = "{A dusty streamer infalling onto the disk of a class I protostar. ALMA dual-band constraints on grain properties and the mass-infall rate}",
      journal = {\aap},
         year = 2024,
        month = feb,
       volume = {682},
          eid = {A61},
        pages = {A61},
          doi = {10.1051/0004-6361/202347486},
archivePrefix = {arXiv},
       eprint = {2311.13723},
 primaryClass = {astro-ph.EP},
       adsurl = {https://ui.adsabs.harvard.edu/abs/2024A&A...682A..61C}
}

@ARTICLE{Hales2024,
       author = {{Hales}, A.~S. and {Gupta}, A. and {Ru{\'\i}z-Rodr{\'\i}guez}, D. and {Williams}, J.~P. and {P{\'e}rez}, S. and {Cieza}, L. and {Gonz{\'a}lez-Ruilova}, C. and {Pineda}, J.~E. and {Santamar{\'\i}a-Miranda}, A. and {Tobin}, J. and {Weber}, P. and {Zhu}, Z. and {Zurlo}, A.},
        title = "{Discovery of an Accretion Streamer and a Slow Wide-angle Outflow around FU Orionis}",
      journal = {\apj},
         year = 2024,
        month = may,
       volume = {966},
       number = {1},
          eid = {96},
        pages = {96},
          doi = {10.3847/1538-4357/ad31a1},
archivePrefix = {arXiv},
       eprint = {2405.03033},
 primaryClass = {astro-ph.SR},
       adsurl = {https://ui.adsabs.harvard.edu/abs/2024ApJ...966...96H}
}

@ARTICLE{Flores2023,
       author = {{Flores}, Christian and {Ohashi}, Nagayoshi and {Tobin}, John J. and {J{\o}rgensen}, Jes K. and {Takakuwa}, Shigehisa and {Li}, Zhi-Yun and {Lin}, Zhe-Yu Daniel and {van't Hoff}, Merel L.~R. and {Plunkett}, Adele L. and {Yamato}, Yoshihide and {Sai (Insa Choi)}, Jinshi and {Koch}, Patrick M. and {Yen}, Hsi-Wei and {Aikawa}, Yuri and {Aso}, Yusuke and {de Gregorio-Monsalvo}, Itziar and {Kido}, Miyu and {Kwon}, Woojin and {Lee}, Jeong-Eun and {Lee}, Chang Won and {Looney}, Leslie W. and {Santamar{\'\i}a-Miranda}, Alejandro and {Sharma}, Rajeeb and {Thieme}, Travis J. and {Williams}, Jonathan P. and {Han}, Ilseung and {Narayanan}, Suchitra and {Lai}, Shih-Ping},
        title = "{Early Planet Formation in Embedded Disks (eDisk). XII. Accretion Streamers, Protoplanetary Disk, and Outflow in the Class I Source Oph IRS 63}",
      journal = {\apj},
         year = 2023,
        month = nov,
       volume = {958},
       number = {1},
          eid = {98},
        pages = {98},
          doi = {10.3847/1538-4357/acf7c1},
archivePrefix = {arXiv},
       eprint = {2310.14617},
 primaryClass = {astro-ph.SR},
       adsurl = {https://ui.adsabs.harvard.edu/abs/2023ApJ...958...98F}
}

@ARTICLE{Trapman2025,
       author = {{Trapman}, Leon and {Zhang}, Ke and {Rosotti}, Giovanni P. and {Pinilla}, Paola and {Tabone}, Beno{\^\i}t and {Pascucci}, Ilaria and {Agurto-Gangas}, Carolina and {Anania}, Rossella and {Carpenter}, John and {Cieza}, Lucas A. and {Deng}, Dingshan and {Gonz{\'a}lez-Ruilova}, Camilo and {Hogerheijde}, Michiel R. and {Kurtovic}, Nicol{\'a}s T. and {Kuznetsova}, Aleksandra and {Miley}, James and {P{\'e}rez}, Laura M. and {Ruiz-Rodriguez}, Dary A. and {Schwarz}, Kamber and {Sierra}, Anibal and {TorresVillanueva}, Estephani and {Vioque}, Miguel},
        title = "{The ALMA Survey of Gas Evolution of PROtoplanetary Disks (AGE-PRO). V. Protoplanetary Gas Disk Masses}",
      journal = {\apj},
         year = 2025,
        month = aug,
       volume = {989},
       number = {1},
          eid = {5},
        pages = {5},
          doi = {10.3847/1538-4357/adcd6e},
archivePrefix = {arXiv},
       eprint = {2506.10738},
 primaryClass = {astro-ph.EP},
       adsurl = {https://ui.adsabs.harvard.edu/abs/2025ApJ...989....5T}
}

@ARTICLE{AgurtoGangas2025,
       author = {{Agurto-Gangas}, Carolina and {P{\'e}rez}, Laura M. and {Sierra}, Anibal and {Miley}, James and {Zhang}, Ke and {Pascucci}, Ilaria and {Pinilla}, Paola and {Deng}, Dingshan and {Carpenter}, John and {Trapman}, Leon and {Vioque}, Miguel and {Rosotti}, Giovanni P. and {Kurtovic}, Nicolas and {Cieza}, Lucas A. and {Anania}, Rossella and {Tabone}, Beno{\^\i}t and {Schwarz}, Kamber and {Hogerheijde}, Michiel R. and {TorresVillanueva}, Estephani E. and {Ruiz-Rodriguez}, Dary A. and {Gonz{\'a}lez-Ruilova}, Camilo},
        title = "{The ALMA Survey of Gas Evolution of PROtoplanetary Disks (AGE-PRO). IV. Dust and Gas Disk Properties in the Upper Scorpius Star-forming Region}",
      journal = {\apj},
         year = 2025,
        month = aug,
       volume = {989},
       number = {1},
          eid = {4},
        pages = {4},
          doi = {10.3847/1538-4357/adc7ab},
archivePrefix = {arXiv},
       eprint = {2506.10735},
 primaryClass = {astro-ph.EP},
       adsurl = {https://ui.adsabs.harvard.edu/abs/2025ApJ...989....4A}
}

@ARTICLE{Deng2025,
       author = {{Deng}, Dingshan and {Vioque}, Miguel and {Pascucci}, Ilaria and {P{\'e}rez}, Laura M. and {Zhang}, Ke and {Kurtovic}, Nicol{\'a}s and {Trapman}, Leon and {TorresVillanueva}, Estephani E. and {Agurto-Gangas}, Carolina and {Carpenter}, John and {Pinilla}, Paola and {Gorti}, Uma and {Tabone}, Beno{\^\i}t and {Sierra}, Anibal and {Rosotti}, Giovanni P. and {Cieza}, Lucas A. and {Anania}, Rossella and {Gonz{\'a}lez-Ruilova}, Camilo and {Hogerheijde}, Michiel R. and {Miley}, James and {Ruiz-Rodriguez}, Dary A. and {Ruaud}, Maxime and {Schwarz}, Kamber},
        title = "{The ALMA Survey of Gas Evolution of PROtoplanetary Disks (AGE-PRO). III. Dust and Gas Disk Properties in the Lupus Star-forming Region}",
      journal = {\apj},
         year = 2025,
        month = aug,
       volume = {989},
       number = {1},
          eid = {3},
        pages = {3},
          doi = {10.3847/1538-4357/add43a},
archivePrefix = {arXiv},
       eprint = {2506.10734},
 primaryClass = {astro-ph.EP},
       adsurl = {https://ui.adsabs.harvard.edu/abs/2025ApJ...989....3D}
}

@ARTICLE{Zhang2025,
       author = {{Zhang}, Ke and {P{\'e}rez}, Laura M. and {Pascucci}, Ilaria and {Pinilla}, Paola and {Cieza}, Lucas A. and {Carpenter}, John and {Trapman}, Leon and {Deng}, Dingshan and {Agurto-Gangas}, Carolina and {Sierra}, Anibal and {Kurtovic}, Nicol{\'a}s T. and {Ruiz-Rodriguez}, Dary A. and {Vioque}, Miguel and {Miley}, James and {Tabone}, Beno{\^\i}t and {Gonz{\'a}lez-Ruilova}, Camilo and {Anania}, Rossella and {Rosotti}, Giovanni P. and {TorresVillanueva}, Estephani and {Hogerheijde}, Michiel R. and {Schwarz}, Kamber and {Kuznetsova}, Aleksandra},
        title = "{The ALMA Survey of Gas Evolution of PROtoplanetary Disks (AGE-PRO). I. Program Overview and Summary of First Results}",
      journal = {\apj},
         year = 2025,
        month = aug,
       volume = {989},
       number = {1},
          eid = {1},
        pages = {1},
          doi = {10.3847/1538-4357/addebe},
archivePrefix = {arXiv},
       eprint = {2506.10719},
 primaryClass = {astro-ph.EP},
       adsurl = {https://ui.adsabs.harvard.edu/abs/2025ApJ...989....1Z}
}

@ARTICLE{Oberg2021,
       author = {{{\"O}berg}, Karin I. and {Guzm{\'a}n}, Viviana V. and {Walsh}, Catherine and {Aikawa}, Yuri and {Bergin}, Edwin A. and {Law}, Charles J. and {Loomis}, Ryan A. and {Alarc{\'o}n}, Felipe and {Andrews}, Sean M. and {Bae}, Jaehan and {Bergner}, Jennifer B. and {Boehler}, Yann and {Booth}, Alice S. and {Bosman}, Arthur D. and {Calahan}, Jenny K. and {Cataldi}, Gianni and {Cleeves}, L. Ilsedore and {Czekala}, Ian and {Furuya}, Kenji and {Huang}, Jane and {Ilee}, John D. and {Kurtovic}, Nicolas T. and {Le Gal}, Romane and {Liu}, Yao and {Long}, Feng and {M{\'e}nard}, Fran{\c{c}}ois and {Nomura}, Hideko and {P{\'e}rez}, Laura M. and {Qi}, Chunhua and {Schwarz}, Kamber R. and {Sierra}, Anibal and {Teague}, Richard and {Tsukagoshi}, Takashi and {Yamato}, Yoshihide and {van't Hoff}, Merel L.~R. and {Waggoner}, Abygail R. and {Wilner}, David J. and {Zhang}, Ke},
        title = "{Molecules with ALMA at Planet-forming Scales (MAPS). I. Program Overview and Highlights}",
      journal = {\apjs},
         year = 2021,
        month = nov,
       volume = {257},
       number = {1},
          eid = {1},
        pages = {1},
          doi = {10.3847/1538-4365/ac1432},
archivePrefix = {arXiv},
       eprint = {2109.06268},
 primaryClass = {astro-ph.EP},
       adsurl = {https://ui.adsabs.harvard.edu/abs/2021ApJS..257....1O}
}

@ARTICLE{Gupta2026,
       author = {{Gupta}, Aashish and {Hales}, Antonio S. and {Cleeves}, L. Ilsedore and {Alves}, Felipe and {Bhowmik}, Trisha and {Cuello}, Nicol{\'a}s and {Girart}, Josep M. and {Li}, Zhi-Yun and {Miotello}, Anna and {Zhu}, Zhaohuan and {Zurlo}, Alice},
        title = "{A Tale of Three Tails: A Misaligned Streamer and Mysterious Structures around [BHB2007]-1}",
      journal = {\apj},
         year = 2026,
        month = feb,
       volume = {998},
       number = {1},
          eid = {105},
        pages = {105},
          doi = {10.3847/1538-4357/ae2477},
archivePrefix = {arXiv},
       eprint = {2512.00295},
 primaryClass = {astro-ph.SR},
       adsurl = {https://ui.adsabs.harvard.edu/abs/2026ApJ...998..105G}
}

@ARTICLE{Mendoza2009,
       author = {{Mendoza}, S. and {Tejeda}, E. and {Nagel}, E.},
        title = "{Analytic solutions to the accretion of a rotating finite cloud towards a central object - I. Newtonian approach}",
      journal = {\mnras},
         year = 2009,
        month = feb,
       volume = {393},
       number = {2},
        pages = {579-586},
          doi = {10.1111/j.1365-2966.2008.14210.x},
archivePrefix = {arXiv},
       eprint = {0803.1020},
 primaryClass = {astro-ph},
       adsurl = {https://ui.adsabs.harvard.edu/abs/2009MNRAS.393..579M}
}

@ARTICLE{Pineda2020,
       author = {{Pineda}, Jaime E. and {Segura-Cox}, Dominique and {Caselli}, Paola and {Cunningham}, Nichol and {Zhao}, Bo and {Schmiedeke}, Anika and {Maureira}, Mar{\'\i}a Jos{\'e} and {Neri}, Roberto},
        title = "{A protostellar system fed by a streamer of 10,500 au length}",
      journal = {Nature Astronomy},
         year = 2020,
        month = jan,
       volume = {4},
        pages = {1158-1163},
          doi = {10.1038/s41550-020-1150-z},
archivePrefix = {arXiv},
       eprint = {2007.13430},
 primaryClass = {astro-ph.GA},
       adsurl = {https://ui.adsabs.harvard.edu/abs/2020NatAs...4.1158P}
}

@ARTICLE{Speedie2025,
       author = {{Speedie}, Jessica and {Dong}, Ruobing and {Teague}, Richard and {Segura-Cox}, Dominique and {Pineda}, Jaime E. and {Calcino}, Josh and {Longarini}, Cristiano and {Hall}, Cassandra and {Tang}, Ya-Wen and {Hashimoto}, Jun and {Paneque-Carre{\~n}o}, Teresa and {Lodato}, Giuseppe and {Veronesi}, Bennedetta},
        title = "{Mapping the Merging Zone of Late Infall in the AB Aur Planet-forming System}",
      journal = {\apjl},
         year = 2025,
        month = mar,
       volume = {981},
       number = {2},
          eid = {L30},
        pages = {L30},
          doi = {10.3847/2041-8213/adb7d5},
archivePrefix = {arXiv},
       eprint = {2503.01957},
 primaryClass = {astro-ph.EP},
       adsurl = {https://ui.adsabs.harvard.edu/abs/2025ApJ...981L..30S}
}

@ARTICLE{Gupta2024,
       author = {{Gupta}, Aashish and {Miotello}, Anna and {Williams}, Jonathan P. and {Birnstiel}, Til and {Kuffmeier}, Michael and {Yen}, Hsi-Wei},
        title = "{TIPSY: Trajectory of Infalling Particles in Streamers around Young stars. Dynamical analysis of the streamers around S CrA and HL Tau}",
      journal = {\aap},
         year = 2024,
        month = mar,
       volume = {683},
          eid = {A133},
        pages = {A133},
          doi = {10.1051/0004-6361/202348007},
archivePrefix = {arXiv},
       eprint = {2401.10403},
 primaryClass = {astro-ph.SR},
       adsurl = {https://ui.adsabs.harvard.edu/abs/2024A&A...683A.133G}
}

@ARTICLE{Bondi1944,
       author = {{Bondi}, H. and {Hoyle}, F.},
        title = "{On the mechanism of accretion by stars}",
      journal = {\mnras},
         year = 1944,
        month = jan,
       volume = {104},
        pages = {273},
          doi = {10.1093/mnras/104.5.273},
       adsurl = {https://ui.adsabs.harvard.edu/abs/1944MNRAS.104..273B}
}

@ARTICLE{Bondi1952,
       author = {{Bondi}, H.},
        title = "{On spherically symmetrical accretion}",
      journal = {\mnras},
         year = 1952,
        month = jan,
       volume = {112},
        pages = {195},
          doi = {10.1093/mnras/112.2.195},
       adsurl = {https://ui.adsabs.harvard.edu/abs/1952MNRAS.112..195B}
}

@ARTICLE{Izquierdo2025,
       author = {{Izquierdo}, Andr{\'e}s F. and {Stadler}, Jochen and {Galloway-Sprietsma}, Maria and {Benisty}, Myriam and {Pinte}, Christophe and {Bae}, Jaehan and {Teague}, Richard and {Facchini}, Stefano and {W{\"o}lfer}, Lisa and {Longarini}, Cristiano and {Curone}, Pietro and {Andrews}, Sean M. and {Barraza-Alfaro}, Marcelo and {Cataldi}, Gianni and {Cuello}, Nicol{\'a}s and {Czekala}, Ian and {Fasano}, Daniele and {Flock}, Mario and {Fukagawa}, Misato and {Garg}, Himanshi and {Hall}, Cassandra and {Hammond}, Iain and {Hilder}, Thomas and {Huang}, Jane and {Ilee}, John D. and {Isella}, Andrea and {Kanagawa}, Kazuhiro and {Lesur}, Geoffroy and {Lodato}, Giuseppe and {Loomis}, Ryan A. and {Orihara}, Ryuta and {Price}, Daniel J. and {Rosotti}, Giovanni and {Testi}, Leonardo and {Yen}, Hsi-Wei and {Wafflard-Fernandez}, Gaylor and {Wilner}, David J. and {Winter}, Andrew J. and {Yoshida}, Tomohiro C. and {Zawadzki}, Brianna},
        title = "{exoALMA. III. Line-intensity Modeling and System Property Extraction from Protoplanetary Disks}",
      journal = {\apjl},
         year = 2025,
        month = may,
       volume = {984},
       number = {1},
          eid = {L8},
        pages = {L8},
          doi = {10.3847/2041-8213/adc439},
archivePrefix = {arXiv},
       eprint = {2504.19986},
 primaryClass = {astro-ph.EP},
       adsurl = {https://ui.adsabs.harvard.edu/abs/2025ApJ...984L...8I}
}

@ARTICLE{Stadler2025,
       author = {{Stadler}, Jochen and {Benisty}, Myriam and {Winter}, Andrew J. and {Izquierdo}, Andr{\'e}s F. and {Longarini}, Cristiano and {Galloway-Sprietsma}, Maria and {Curone}, Pietro and {Andrews}, Sean M. and {Bae}, Jaehan and {Facchini}, Stefano and {Rosotti}, Giovanni and {Teague}, Richard and {Barraza-Alfaro}, Marcelo and {Cataldi}, Gianni and {Cuello}, Nicol{\'a}s and {Czekala}, Ian and {Fasano}, Daniele and {Flock}, Mario and {Fukagawa}, Misato and {Garg}, Himanshi and {Hall}, Cassandra and {Hammond}, Iain and {Hilder}, Thomas and {Huang}, Jane and {Ilee}, John D. and {Kanagawa}, Kazuhiro and {Lesur}, Geoffroy and {Lodato}, Giuseppe and {Loomis}, Ryan A. and {Menard}, Francois and {Orihara}, Ryuta and {Pinte}, Christophe and {Price}, Daniel J. and {Yen}, Hsi-Wei and {Wafflard-Fernandez}, Gaylor and {Wilner}, David J. and {W{\"o}lfer}, Lisa and {Yoshida}, Tomohiro C. and {Zawadzki}, Brianna},
        title = "{exoALMA. VI. Rotating under Pressure: Rotation Curves, Azimuthal Velocity Substructures, and Gas Pressure Variations}",
      journal = {\apjl},
         year = 2025,
        month = may,
       volume = {984},
       number = {1},
          eid = {L11},
        pages = {L11},
          doi = {10.3847/2041-8213/adb152},
archivePrefix = {arXiv},
       eprint = {2504.20036},
 primaryClass = {astro-ph.EP},
       adsurl = {https://ui.adsabs.harvard.edu/abs/2025ApJ...984L..11S}
}

@ARTICLE{Martire2024,
       author = {{Martire}, P. and {Longarini}, C. and {Lodato}, G. and {Rosotti}, G.~P. and {Winter}, A. and {Facchini}, S. and {Hardiman}, C. and {Benisty}, M. and {Stadler}, J. and {Izquierdo}, A.~F. and {Testi}, Leonardo},
        title = "{Rotation curves in protoplanetary disks with thermal stratification. Physical model and observational evidence in MAPS disks}",
      journal = {\aap},
         year = 2024,
        month = jun,
       volume = {686},
          eid = {A9},
        pages = {A9},
          doi = {10.1051/0004-6361/202348546},
archivePrefix = {arXiv},
       eprint = {2402.12236},
 primaryClass = {astro-ph.EP},
       adsurl = {https://ui.adsabs.harvard.edu/abs/2024A&A...686A...9M}
}

@ARTICLE{LyndenBell1974,
       author = {{Lynden-Bell}, D. and {Pringle}, J.~E.},
        title = "{The evolution of viscous discs and the origin of the nebular variables.}",
      journal = {\mnras},
         year = 1974,
        month = sep,
       volume = {168},
        pages = {603-637},
          doi = {10.1093/mnras/168.3.603},
       adsurl = {https://ui.adsabs.harvard.edu/abs/1974MNRAS.168..603L}
}

@ARTICLE{GallowaySprietsma2025,
       author = {{Galloway-Sprietsma}, Maria and {Bae}, Jaehan and {Izquierdo}, Andr{\'e}s F. and {Stadler}, Jochen and {Longarini}, Cristiano and {Teague}, Richard and {Andrews}, Sean M. and {Winter}, Andrew J. and {Benisty}, Myriam and {Facchini}, Stefano and {Rosotti}, Giovanni and {Zawadzki}, Brianna and {Pinte}, Christophe and {Fasano}, Daniele and {Barraza-Alfaro}, Marcelo and {Cataldi}, Gianni and {Cuello}, Nicol{\'a}s and {Curone}, Pietro and {Czekala}, Ian and {Flock}, Mario and {Fukagawa}, Misato and {Gardner}, Charles H. and {Garg}, Himanshi and {Hall}, Cassandra and {Huang}, Jane and {Ilee}, John D. and {Kanagawa}, Kazuhiro and {Lesur}, Geoffroy and {Lodato}, Giuseppe and {Loomis}, Ryan A. and {Menard}, Francois and {Orihara}, Ryuta and {Price}, Daniel J. and {Wafflard-Fernandez}, Gaylor and {Wilner}, David J. and {W{\"o}lfer}, Lisa and {Yen}, Hsi-Wei and {Yoshida}, Tomohiro C.},
        title = "{exoALMA. V. Gaseous Emission Surfaces and Temperature Structures}",
      journal = {\apjl},
         year = 2025,
        month = may,
       volume = {984},
       number = {1},
          eid = {L10},
        pages = {L10},
          doi = {10.3847/2041-8213/adc437},
archivePrefix = {arXiv},
       eprint = {2504.19902},
 primaryClass = {astro-ph.EP},
       adsurl = {https://ui.adsabs.harvard.edu/abs/2025ApJ...984L..10G}
}

@ARTICLE{Teague2025,
       author = {{Teague}, Richard and {Benisty}, Myriam and {Facchini}, Stefano and {Fukagawa}, Misato and {Pinte}, Christophe and {Andrews}, Sean M. and {Bae}, Jaehan and {Barraza-Alfaro}, Marcelo and {Cataldi}, Gianni and {Cuello}, Nicol{\'a}s and {Curone}, Pietro and {Czekala}, Ian and {Fasano}, Daniele and {Flock}, Mario and {Galloway-Sprietsma}, Maria and {Garg}, Himanshi and {Hall}, Cassandra and {Hammond}, Iain and {Hilder}, Thomas and {Huang}, Jane and {Ilee}, John D. and {Izquierdo}, Andr{\'e}s F. and {Kanagawa}, Kazuhiro and {Lesur}, Geoffroy and {Lodato}, Giuseppe and {Longarini}, Cristiano and {Loomis}, Ryan A. and {Masset}, Fr{\'e}d{\'e}ric and {Menard}, Francois and {Orihara}, Ryuta and {Price}, Daniel J. and {Rosotti}, Giovanni and {Stadler}, Jochen and {Testi}, Leonardo and {Yen}, Hsi-Wei and {Wafflard-Fernandez}, Gaylor and {Wilner}, David J. and {Winter}, Andrew J. and {W{\"o}lfer}, Lisa and {Yoshida}, Tomohiro C. and {Zawadzki}, Brianna},
        title = "{exoALMA. I. Science Goals, Project Design, and Data Products}",
      journal = {\apjl},
         year = 2025,
        month = may,
       volume = {984},
       number = {1},
          eid = {L6},
        pages = {L6},
          doi = {10.3847/2041-8213/adc43b},
archivePrefix = {arXiv},
       eprint = {2504.18688},
 primaryClass = {astro-ph.EP},
       adsurl = {https://ui.adsabs.harvard.edu/abs/2025ApJ...984L...6T}
}

@ARTICLE{Longarini2025,
       author = {{Longarini}, Cristiano and {Lodato}, Giuseppe and {Rosotti}, Giovanni and {Andrews}, Sean and {Winter}, Andrew and {Stadler}, Jochen and {Izquierdo}, Andr{\'e}s and {Galloway-Sprietsma}, Maria and {Facchini}, Stefano and {Curone}, Pietro and {Benisty}, Myriam and {Teague}, Richard and {Bae}, Jaehan and {Barraza-Alfaro}, Marcelo and {Cataldi}, Gianni and {Czekala}, Ian and {Cuello}, Nicol{\'a}s and {Fasano}, Daniele and {Flock}, Mario and {Fukagawa}, Misato and {Garg}, Himanshi and {Hall}, Cassandra and {Hammond}, Iain and {Hardiman}, Caitlyn and {Hilder}, Thomas and {Huang}, Jane and {Ilee}, John D. and {Isella}, Andrea and {Kanagawa}, Kazuhiro and {Lesur}, Geoffroy and {Loomis}, Ryan A. and {M{\'e}nard}, Francois and {Orihara}, Ryuta and {Pinte}, Christophe and {Price}, Daniel and {Testi}, Leonardo and {Fernandez}, Gaylor Wafflard- and {W{\"o}lfer}, Lisa and {Yen}, Hsi-Wei and {Yoshida}, Tomohiro C. and {Zawadzki}, Brianna},
        title = "{exoALMA. XII. Weighing and Sizing exoALMA Disks with Rotation Curve Modelling}",
      journal = {\apjl},
         year = 2025,
        month = may,
       volume = {984},
       number = {1},
          eid = {L17},
        pages = {L17},
          doi = {10.3847/2041-8213/adc431},
archivePrefix = {arXiv},
       eprint = {2504.18726},
 primaryClass = {astro-ph.EP},
       adsurl = {https://ui.adsabs.harvard.edu/abs/2025ApJ...984L..17L}
}

@ARTICLE{Tanious2024,
       author = {{Tanious}, M. and {Le Gal}, R. and {Neri}, R. and {Faure}, A. and {Gupta}, A. and {Law}, C.~J. and {Huang}, J. and {Cuello}, N. and {Williams}, J.~P. and {M{\'e}nard}, F.},
        title = "{Anatomy of the Class I protostar L1489 IRS with NOEMA. I. Disk, streamers, outflow(s) and bubbles at 3 mm}",
      journal = {\aap},
         year = 2024,
        month = jul,
       volume = {687},
          eid = {A92},
        pages = {A92},
          doi = {10.1051/0004-6361/202348785},
archivePrefix = {arXiv},
       eprint = {2403.18905},
 primaryClass = {astro-ph.SR},
       adsurl = {https://ui.adsabs.harvard.edu/abs/2024A&A...687A..92T}
}

@ARTICLE{Pelkonen2025,
       author = {{Pelkonen}, V. -M. and {Padoan}, P. and {Juvela}, M. and {Haugb{\o}lle}, T. and {Nordlund}, {\r{A}}.},
        title = "{Origin and evolution of angular momentum of class II disks}",
      journal = {\aap},
         year = 2025,
        month = feb,
       volume = {694},
          eid = {A327},
        pages = {A327},
          doi = {10.1051/0004-6361/202450682},
archivePrefix = {arXiv},
       eprint = {2405.06520},
 primaryClass = {astro-ph.SR},
       adsurl = {https://ui.adsabs.harvard.edu/abs/2025A&A...694A.327P}
}

@ARTICLE{Kuffmeier2024,
       author = {{Kuffmeier}, M. and {Pineda}, J.~E. and {Segura-Cox}, D. and {Haugb{\o}lle}, T.},
        title = "{Constraints on the primordial misalignment of star-disk systems}",
      journal = {\aap},
         year = 2024,
        month = oct,
       volume = {690},
          eid = {A297},
        pages = {A297},
          doi = {10.1051/0004-6361/202450410},
archivePrefix = {arXiv},
       eprint = {2405.12670},
 primaryClass = {astro-ph.SR},
       adsurl = {https://ui.adsabs.harvard.edu/abs/2024A&A...690A.297K}
}

@ARTICLE{Kuffmeier2021,
       author = {{Kuffmeier}, M. and {Dullemond}, C.~P. and {Reissl}, S. and {Goicovic}, F.~G.},
        title = "{Misaligned disks induced by infall}",
      journal = {\aap},
         year = 2021,
        month = dec,
       volume = {656},
          eid = {A161},
        pages = {A161},
          doi = {10.1051/0004-6361/202039614},
archivePrefix = {arXiv},
       eprint = {2110.04309},
 primaryClass = {astro-ph.SR},
       adsurl = {https://ui.adsabs.harvard.edu/abs/2021A&A...656A.161K}
}

@ARTICLE{Padoan2025,
       author = {{Padoan}, Paolo and {Pan}, Liubin and {Pelkonen}, Veli-Matti and {Haugb{\o}lle}, Troels and {Nordlund}, {\^a}.{\guillemotleft}ke},
        title = "{The formation of protoplanetary disks through pre-main-sequence Bondi-Hoyle accretion}",
      journal = {Nature Astronomy},
         year = 2025,
        month = jun,
       volume = {9},
        pages = {862-871},
          doi = {10.1038/s41550-025-02529-3},
archivePrefix = {arXiv},
       eprint = {2405.07334},
 primaryClass = {astro-ph.GA},
       adsurl = {https://ui.adsabs.harvard.edu/abs/2025NatAs...9..862P}
}

@ARTICLE{Thies2011,
       author = {{Thies}, I. and {Kroupa}, P. and {Goodwin}, S.~P. and {Stamatellos}, D. and {Whitworth}, A.~P.},
        title = "{A natural formation scenario for misaligned and short-period eccentric extrasolar planets}",
      journal = {\mnras},
         year = 2011,
        month = nov,
       volume = {417},
       number = {3},
        pages = {1817-1822},
          doi = {10.1111/j.1365-2966.2011.19390.x},
archivePrefix = {arXiv},
       eprint = {1107.2113},
 primaryClass = {astro-ph.EP},
       adsurl = {https://ui.adsabs.harvard.edu/abs/2011MNRAS.417.1817T}
}

@ARTICLE{Nealon2020,
       author = {{Nealon}, Rebecca and {Cuello}, Nicol{\'a}s and {Alexander}, Richard},
        title = "{Flyby-induced misalignments in planet-hosting discs}",
      journal = {\mnras},
         year = 2020,
        month = jan,
       volume = {491},
       number = {3},
        pages = {4108-4115},
          doi = {10.1093/mnras/stz3186},
archivePrefix = {arXiv},
       eprint = {1911.05760},
 primaryClass = {astro-ph.EP},
       adsurl = {https://ui.adsabs.harvard.edu/abs/2020MNRAS.491.4108N}
}

@ARTICLE{Dullemond2019,
       author = {{Dullemond}, C.~P. and {K{\"u}ffmeier}, M. and {Goicovic}, F. and {Fukagawa}, M. and {Oehl}, V. and {Kramer}, M.},
        title = "{Cloudlet capture by transitional disk and FU Orionis stars}",
      journal = {\aap},
         year = 2019,
        month = aug,
       volume = {628},
          eid = {A20},
        pages = {A20},
          doi = {10.1051/0004-6361/201832632},
archivePrefix = {arXiv},
       eprint = {1911.05158},
 primaryClass = {astro-ph.EP},
       adsurl = {https://ui.adsabs.harvard.edu/abs/2019A&A...628A..20D}
}

@ARTICLE{Cuello2023,
       author = {{Cuello}, Nicol{\'a}s and {M{\'e}nard}, Fran{\c{c}}ois and {Price}, Daniel J.},
        title = "{Close encounters: How stellar flybys shape planet-forming discs}",
      journal = {European Physical Journal Plus},
         year = 2023,
        month = jan,
       volume = {138},
       number = {1},
          eid = {11},
        pages = {11},
          doi = {10.1140/epjp/s13360-022-03602-w},
archivePrefix = {arXiv},
       eprint = {2207.09752},
 primaryClass = {astro-ph.EP},
       adsurl = {https://ui.adsabs.harvard.edu/abs/2023EPJP..138...11C}
}

@ARTICLE{Luhman2020,
       author = {{Luhman}, K.~L.},
        title = "{A Gaia Survey for Young Stars Associated with the Lupus Clouds}",
      journal = {\aj},
         year = 2020,
        month = oct,
       volume = {160},
       number = {4},
          eid = {186},
        pages = {186},
          doi = {10.3847/1538-3881/abb12f},
archivePrefix = {arXiv},
       eprint = {2009.05123},
 primaryClass = {astro-ph.SR},
       adsurl = {https://ui.adsabs.harvard.edu/abs/2020AJ....160..186L}
}

@ARTICLE{Gupta2022,
       author = {{Gupta}, Aashish and {Chen}, Wen-Ping},
        title = "{Interplay between Young Stars and Molecular Clouds in the Ophiuchus Star-forming Complex}",
      journal = {\aj},
         year = 2022,
        month = may,
       volume = {163},
       number = {5},
          eid = {233},
        pages = {233},
          doi = {10.3847/1538-3881/ac5cc8},
archivePrefix = {arXiv},
       eprint = {2204.13797},
 primaryClass = {astro-ph.SR},
       adsurl = {https://ui.adsabs.harvard.edu/abs/2022AJ....163..233G}
}

@ARTICLE{Adams2010,
       author = {{Adams}, Fred C.},
        title = "{The Birth Environment of the Solar System}",
      journal = {\araa},
         year = 2010,
        month = sep,
       volume = {48},
        pages = {47-85},
          doi = {10.1146/annurev-astro-081309-130830},
archivePrefix = {arXiv},
       eprint = {1001.5444},
 primaryClass = {astro-ph.SR},
       adsurl = {https://ui.adsabs.harvard.edu/abs/2010ARA&A..48...47A}
}

\begin{appendix}

\section{Other implications of disk angular momentum} \label{app:other}

Angular momentum is a fundamental property of disks and can therefore be used to investigate disk-related processes beyond misalignment. An important open question in disk formation is how much angular momentum is lost due to magnetic braking when pre-stellar cores collapse to form protostars and disks \citep[e.g.,][]{all03,gal06,mel08,joo12,Gupta2022_mag}. While specific angular momentum profiles in prestellar cores and infalling envelopes have been extensively studied \citep[e.g.,][]{Ohashi1997,Belloche2002,Lee2016,Yen2017,Pineda2019,Gaudel2020}, these studies typically focus on individual systems. Comparing angular momentum differences across populations of pre-stellar cores and disks requires total angular momentum values, which depend on mass estimates. The sample presented here offers an opportunity to make this population-level comparison. 

\begin{figure}[htbp]
    \centering
    \begin{subfigure}{0.4\textwidth}
        \centering
        \includegraphics[width=0.9\linewidth]{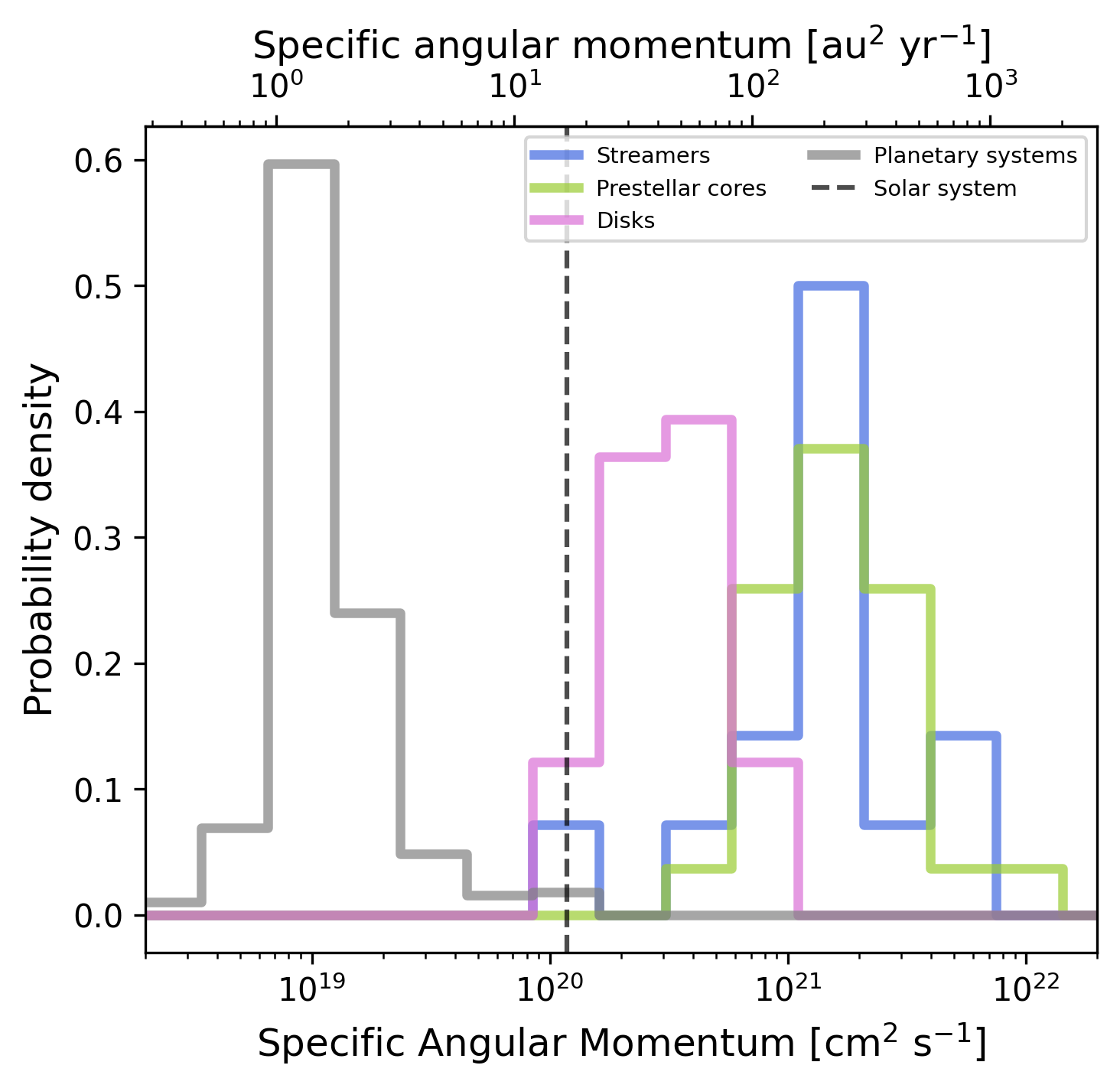}
    \end{subfigure}
    \begin{subfigure}{0.4\textwidth}
        \centering
        \includegraphics[width=0.9\linewidth]{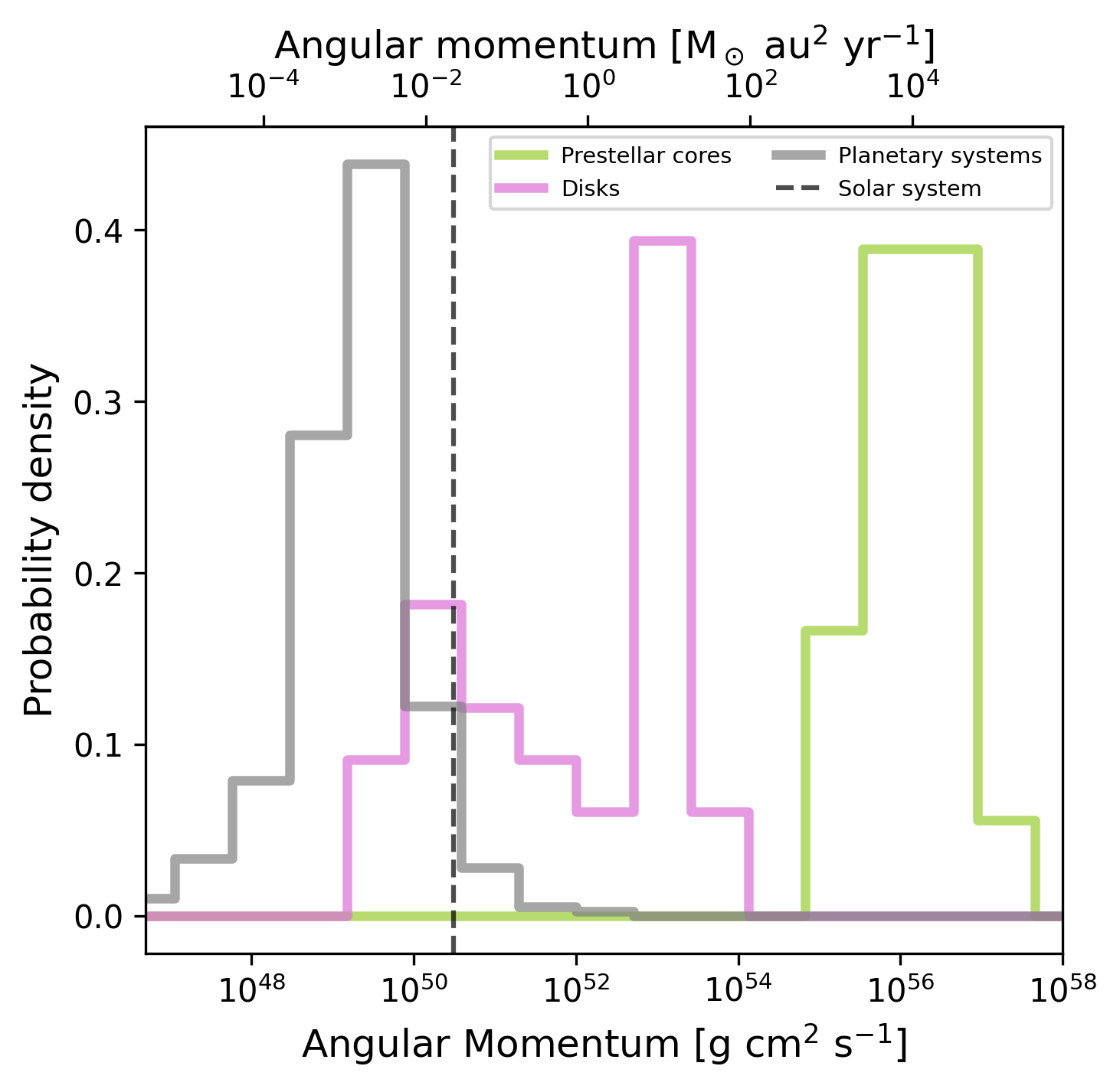}
    \end{subfigure}
    \caption{Specific angular momentum (top panel) and total angular momentum (bottom panel) distributions for streamers from Table \ref{tab:streamers} (blue), disks from Table \ref{tab:disk_properties} (purple), prestellar cores from \cite{Tatematsu2016}, and planetary systems from \citet{Jiang2022} (grey). Total angular momentum values of streamers are not shown because they are expected to be lower limits, as discussed in Section \ref{sec:infall_streamers}. In both panels, dashed vertical lines represent values for planets in Solar system.}
    \label{fig:histograms}
\end{figure}

The average angular momentum of disks in our sample is $\sim5\times10^{52}$~g~cm$^2$~s$^{-1}$. This is roughly four orders of magnitude smaller than the average angular momentum of prestellar cores ($\sim2\times10^{56}$~g~cm$^2$~s$^{-1}$), as derived using masses and specific angular momentum estimated for 27 cores in Orion A cloud in \citet{Tatematsu2016} (see Figure \ref{fig:histograms}). This suggests that cores lose most of their angular momentum during disk formation.
However, it is unclear if the disks eventually formed by these Orion A cores will be similar to the sample of disks analyzed here, although both samples are biased towards biggest systems. Interestingly, distribution of specific angular momentum of these cores are remarkably similar to those of observed streamers, suggesting that similar processes dominate their dynamics.

Another important problem associated with disks is how they lose angular momentum, resulting in mass accretion onto the central star and eventually giving rise to planetary systems \citep[e.g.,][]{Morbidelli2016,Lesur2021,Tabone2022,Manara2023,Rosotti2023,Somigliana2023}. The total orbital angular momentum of a planetary system provides a lower limit on the angular momentum retained by the disk at the end of its evolution, as planets form from disk material and disks can also disperse without forming planets. As a sanity check, we compare the angular momentum of observed planetary systems with that of the disks in our sample. The average orbital angular momentum of observed planetary systems, as compiled in \citet{Jiang2022}, is $\sim5\times10^{49}$~g~cm$^2$~s$^{-1}$, roughly three orders of magnitude lower than the typical disk angular momentum in our sample (see Figure \ref{fig:histograms}), consistent with the expectation that disks lose most of their angular momentum before dispersal. However, the sample of planets is mostly detected via the transit method \citep[e.g.,][]{Borucki2010}, biasing it towards planets closer to their host stars and therefore towards lower angular momentum. This is likely why the solar system, with an angular momentum of $\sim3\times10^{50}$~g~cm$^2$~s$^{-1}$, lies at the higher end of this distribution. Future planet detection statistics, particularly using upcoming data releases from the \textit{Gaia} mission \citep[][]{Gaia2016} and observations from the \textit{Roman Space Telescope}, are expected to reveal planets with larger angular momentum \citep[e.g.,][]{Stefansson2025}, allowing for a more representative comparison with disks. Similarly, upcoming disk surveys such as the Disk-Exoplanet C/Onnection (DEC/O; project id. 2022.1.00875.L) will allow us to estimate the angular momentum of a more representative sample of disks.

\section{Disk properties} \label{app:disks}

Table \ref{tab:disk_properties} presents physical properties of all planet-forming disks for which we estimated angular momentum, as discussed in Section \ref{sec:disks}.

\begin{table*}
\centering
\caption{Disk properties}
\label{tab:disk_properties}
\renewcommand{\arraystretch}{1.3}
\begin{tabular}{lcccccc}
\hline\hline
Source & Sample & $M_{\star}$ [$M_\odot$] & $M_{\rm disk}$ [$M_\odot$] & $R_{90}^{13\rm CO}$ [au] & $L_{\rm tot}$ [g\,cm$^{2}$\,s$^{-1}$] & $j_{\rm tot}$ [cm$^{2}$\,s$^{-1}$] \\
\midrule
AA Tau & exoALMA & $0.6$ & $1.5^{+0.4}_{-0.4}\times10^{-1}$ & $235$ & $1.3^{+0.5}_{-0.4}\times10^{53}$ & $4.4^{+0.9}_{-0.8}\times10^{20}$ \\
MWC 480 & MAPS & $2.0$ & $1.5^{+0.4}_{-0.4}\times10^{-1}$ & $419$ & $1.9^{+0.6}_{-0.5}\times10^{53}$ & $6.5^{+0.7}_{-0.6}\times10^{20}$ \\
HD 34282 & exoALMA & $1.5$ & $1.4^{+0.4}_{-0.4}\times10^{-1}$ & $562$ & $2.7^{+0.9}_{-0.8}\times10^{53}$ & $9.5^{+1.1}_{-1.1}\times10^{20}$ \\
HD 163296 & MAPS & $1.9$ & $1.3^{+0.4}_{-0.3}\times10^{-1}$ & $364$ & $1.4^{+0.4}_{-0.4}\times10^{53}$ & $5.4^{+0.5}_{-0.5}\times10^{20}$ \\
GM Aur & MAPS & $1.1$ & $1.2^{+0.3}_{-0.3}\times10^{-1}$ & $427$ & $9.9^{+2.7}_{-2.5}\times10^{52}$ & $4.2^{+0.4}_{-0.4}\times10^{20}$ \\
LkCa15 & exoALMA & $1.1$ & $1.1^{+0.2}_{-0.2}\times10^{-1}$ & $533$ & $1.1^{+0.2}_{-0.2}\times10^{53}$ & $5.0^{+0.2}_{-0.2}\times10^{20}$ \\
IM Lup & MAPS & $1.2$ & $1.1^{+0.3}_{-0.3}\times10^{-1}$ & $540$ & $9.9^{+3.1}_{-2.3}\times10^{52}$ & $4.7^{+0.5}_{-0.4}\times10^{20}$ \\
SY Cha & exoALMA & $0.8$ & $8.4^{+4.4}_{-4.3}\times10^{-2}$ & $358$ & $6.3^{+3.4}_{-3.0}\times10^{52}$ & $3.8^{+0.3}_{-0.3}\times10^{20}$ \\
RXJ1615 & exoALMA & $1.1$ & $8.2^{+1.4}_{-1.4}\times10^{-2}$ & $348$ & $8.8^{+1.5}_{-1.5}\times10^{52}$ & $5.4^{+0.3}_{-0.3}\times10^{20}$ \\
RXJ1842 & exoALMA & $1.0$ & $7.8^{+1.3}_{-1.4}\times10^{-2}$ & $224$ & $9.9^{+2.2}_{-2.1}\times10^{52}$ & $6.5^{+1.0}_{-0.9}\times10^{20}$ \\
AS 209 & MAPS & $1.3$ & $6.5^{+1.8}_{-1.7}\times10^{-2}$ & $196$ & $6.7^{+2.1}_{-1.7}\times10^{52}$ & $5.2^{+0.5}_{-0.5}\times10^{20}$ \\
V4046 Sgr & exoALMA & $1.8$ & $5.8^{+0.6}_{-0.6}\times10^{-2}$ & $210$ & $6.0^{+0.7}_{-0.6}\times10^{52}$ & $5.2^{+0.1}_{-0.1}\times10^{20}$ \\
DM Tau & exoALMA & $0.5$ & $5.7^{+1.9}_{-2.0}\times10^{-2}$ & $462$ & $4.8^{+1.7}_{-1.7}\times10^{52}$ & $4.3^{+0.3}_{-0.3}\times10^{20}$ \\
RXJ1852 & exoALMA & $1.0$ & $4.4^{+2.4}_{-3.2}\times10^{-2}$ & $157$ & $3.4^{+2.4}_{-2.0}\times10^{52}$ & $4.4^{+0.9}_{-0.8}\times10^{20}$ \\
PDS 66 & exoALMA & $1.3$ & $3.8^{+9.9}_{-3.5}\times10^{-2}$ & $83$ & $3.8^{+4.2}_{-2.6}\times10^{52}$ & $2.5^{+0.4}_{-0.3}\times10^{20}$ \\
J15392776-3446171 & AGEPRO & $0.7$ & $2.0^{+1.1}_{-0.7}\times10^{-3}$ & $195$ & $1.2^{+0.7}_{-0.4}\times10^{51}$ & $2.9^{+2.5}_{-1.3}\times10^{20}$ \\
J15464473-3430354 & AGEPRO & $0.4$ & $9.0^{+6.5}_{-4.1}\times10^{-3}$ & $228$ & $4.5^{+3.3}_{-2.1}\times10^{51}$ & $2.6^{+3.0}_{-1.5}\times10^{20}$ \\
J16124373-3815031 & AGEPRO & $0.5$ & $1.4^{+1.0}_{-0.6}\times10^{-3}$ & $71$ & $4.3^{+2.9}_{-1.8}\times10^{50}$ & $1.5^{+1.6}_{-0.8}\times10^{20}$ \\
J15475062-3528353 & AGEPRO & $0.4$ & $2.3^{+3.5}_{-1.3}\times10^{-4}$ & $118$ & $8.0^{+12.3}_{-4.2}\times10^{49}$ & $1.8^{+4.0}_{-1.3}\times10^{20}$ \\
J15514695-3556440 & AGEPRO & $0.7$ & $8.0^{+22.6}_{-5.4}\times10^{-5}$ & $55$ & $2.6^{+7.5}_{-1.7}\times10^{49}$ & $1.7^{+7.5}_{-1.4}\times10^{20}$ \\
J16085324-3914401 & AGEPRO & $0.3$ & $6.6^{+8.3}_{-3.5}\times10^{-4}$ & $137$ & $2.2^{+2.7}_{-1.2}\times10^{50}$ & $1.6^{+3.4}_{-1.1}\times10^{20}$ \\
J16004943-4130038 & AGEPRO & $0.3$ & $2.7^{+7.6}_{-1.9}\times10^{-4}$ & $67$ & $6.2^{+17.8}_{-4.3}\times10^{49}$ & $1.1^{+5.4}_{-0.9}\times10^{20}$ \\
J15392828-3446180 & AGEPRO & $0.3$ & $8.3^{+11.6}_{-4.5}\times10^{-4}$ & $170$ & $3.0^{+4.4}_{-1.7}\times10^{50}$ & $1.7^{+3.8}_{-1.1}\times10^{20}$ \\
J16075230-3858059 & AGEPRO & $0.3$ & $4.8^{+26.0}_{-4.0}\times10^{-4}$ & $162$ & $1.7^{+9.1}_{-1.4}\times10^{50}$ & $1.7^{+15.8}_{-1.6}\times10^{20}$ \\
J16083617-3923024 & AGEPRO & $0.8$ & $6.3^{+1.9}_{-1.5}\times10^{-2}$ & $739$ & $7.8^{+2.5}_{-1.9}\times10^{52}$ & $6.3^{+3.1}_{-2.1}\times10^{20}$ \\
J16120668-3010270 & AGEPRO & $0.5$ & $3.4^{+1.4}_{-0.9}\times10^{-3}$ & $154$ & $1.5^{+0.7}_{-0.4}\times10^{51}$ & $2.3^{+1.2}_{-0.8}\times10^{20}$ \\
J16054540-2023088 & AGEPRO & $0.1$ & $1.5^{+1.5}_{-0.7}\times10^{-4}$ & $146$ & $3.3^{+3.3}_{-1.5}\times10^{49}$ & $1.1^{+1.6}_{-0.7}\times10^{20}$ \\
J16020757-2257467 & AGEPRO & $0.4$ & $4.9^{+5.0}_{-2.8}\times10^{-4}$ & $102$ & $1.5^{+1.5}_{-0.9}\times10^{50}$ & $1.6^{+3.0}_{-1.0}\times10^{20}$ \\
J16163345-2521505  & AGEPRO & $0.5$ & $6.9^{+8.8}_{-3.7}\times10^{-5}$ & $290$ & $4.3^{+5.7}_{-2.4}\times10^{49}$ & $3.1^{+6.4}_{-2.0}\times10^{20}$ \\
J16202863-2442087 & AGEPRO & $0.3$ & $4.6^{+3.7}_{-2.0}\times10^{-4}$ & $225$ & $2.0^{+1.8}_{-0.9}\times10^{50}$ & $2.3^{+2.8}_{-1.3}\times10^{20}$ \\
J16221532-2511349 & AGEPRO & $0.3$ & $3.4^{+3.5}_{-1.7}\times10^{-3}$ & $112$ & $9.8^{+10.0}_{-5.1}\times10^{50}$ & $1.4^{+2.7}_{-0.9}\times10^{20}$ \\
J16082324-1930009  & AGEPRO & $0.6$ & $4.0^{+3.4}_{-2.2}\times10^{-2}$ & $220$ & $2.3^{+1.9}_{-1.2}\times10^{52}$ & $2.8^{+4.5}_{-1.7}\times10^{20}$ \\
J16090075-1908526 & AGEPRO & $0.5$ & $7.1^{+11.6}_{-4.4}\times10^{-3}$ & $135$ & $3.1^{+5.3}_{-1.9}\times10^{51}$ & $2.3^{+5.9}_{-1.8}\times10^{20}$ \\
\hline
\end{tabular}
\tablefoot{
Uncertainties on $M_*$ are $\sim2\%$ for exoALMA \citep{Longarini2025} and MAPS \citep{Martire2024} sources, and $\sim20\%$ for AGEPRO sources \citep[][]{Deng2025,AgurtoGangas2025}. Uncertainties on $R_{\rm disk; 13CO}$ are typically $10\%$, however it can approach $\sim50\%$ for some faint disks in AGEPRO. The table is also available in electronic form at the CDS via https://cdsarc.cds.unistra.fr/cgi-bin/qcat?J/A+A/VVV/AXX.
}
\end{table*}

\FloatBarrier

\section{Effects of different $\gamma$ values in surface density profiles} \label{app:gamma}

As discussed in Section \ref{sec:surface_density} and shown in Equation \ref{equ:sdenity}, the surface density profiles on the disks also depend on power-law factor $\gamma$, which is generally assumed to be one, consistent with the primary analysis presented in this manuscript. A higher $\gamma$ would mean that the disk surface density is dropping faster. To understand the impact of this factor on our final angular momentum values, we refitted profiles for the ten exoALMA disks \citep[][]{Longarini2025} by setting $\gamma$ to be 0.5 and 1.5. Among these disks, the fits did not converge well for DM Tau, LkCa15, and PDS 66 for $\gamma=1.5$, therefore, we exclude them from the comparison here. 

Table \ref{tab:gamma_var} lists the mean fractional changes, along corresponding standard deviations represented are uncertainties, in disk properties directly fitted using the procedure described in Section \ref{sec:surface_density} and angular momentum properties of the disks estimated from resulting surface density profiles. 
We see modest changes ($\lesssim10\%$) in the estimated stellar masses ($M_{*}$). The higher $\gamma$ means a lower scale radius ($R_{\rm c}$) because of more compact disks, with typical fractional changes of $\sim70\%$. Interestingly, disk masses ($M_{\rm disk}$) shows the opposite trend, albeit with a large scatter, where an increase in $\gamma$ results in an increase in estimated disk mass. These result in fractional changes in specific angular momentum ($j_{\rm disk}$) and total angular momentum ($L_{\rm disk}$) of few tens of percent for most disks. These are within the typical uncertainties derived on these parameters (see Table \ref{tab:disk_properties}) and suggests that we can still use fits for $\gamma=1$ to get rough estimates on angular momentum.

\begin{table}[htbp]
\caption{Effects of $\gamma$ on disk properties}
\label{tab:gamma_var}
\centering
\begin{tabular}{lcc}
\hline\hline
 & $\gamma=1.5$ & $\gamma=0.5$ \\
\hline
$\Delta M_{*}$ & $-4\pm11\%$ & $2\pm2\%$ \\
$\Delta M_{\rm disk}$ & $7\pm82\%$ & $-38\pm25\%$ \\
$\Delta R_{\rm c}$ & $-68\pm21\%$ & $74\pm33\%$ \\
$\Delta L_{\rm disk}$ & $-57\pm34\%$ & $-20\pm29\%$ \\
$\Delta j_{\rm disk}$ & $-58\pm15\%$ & $35\pm14\%$ \\
\hline
\end{tabular}
\end{table}

\section{Comparison between $^{13}$CO and CO radii} \label{app:radii}

\begin{figure}[htbp]
\centering
\includegraphics[width=0.98\linewidth]{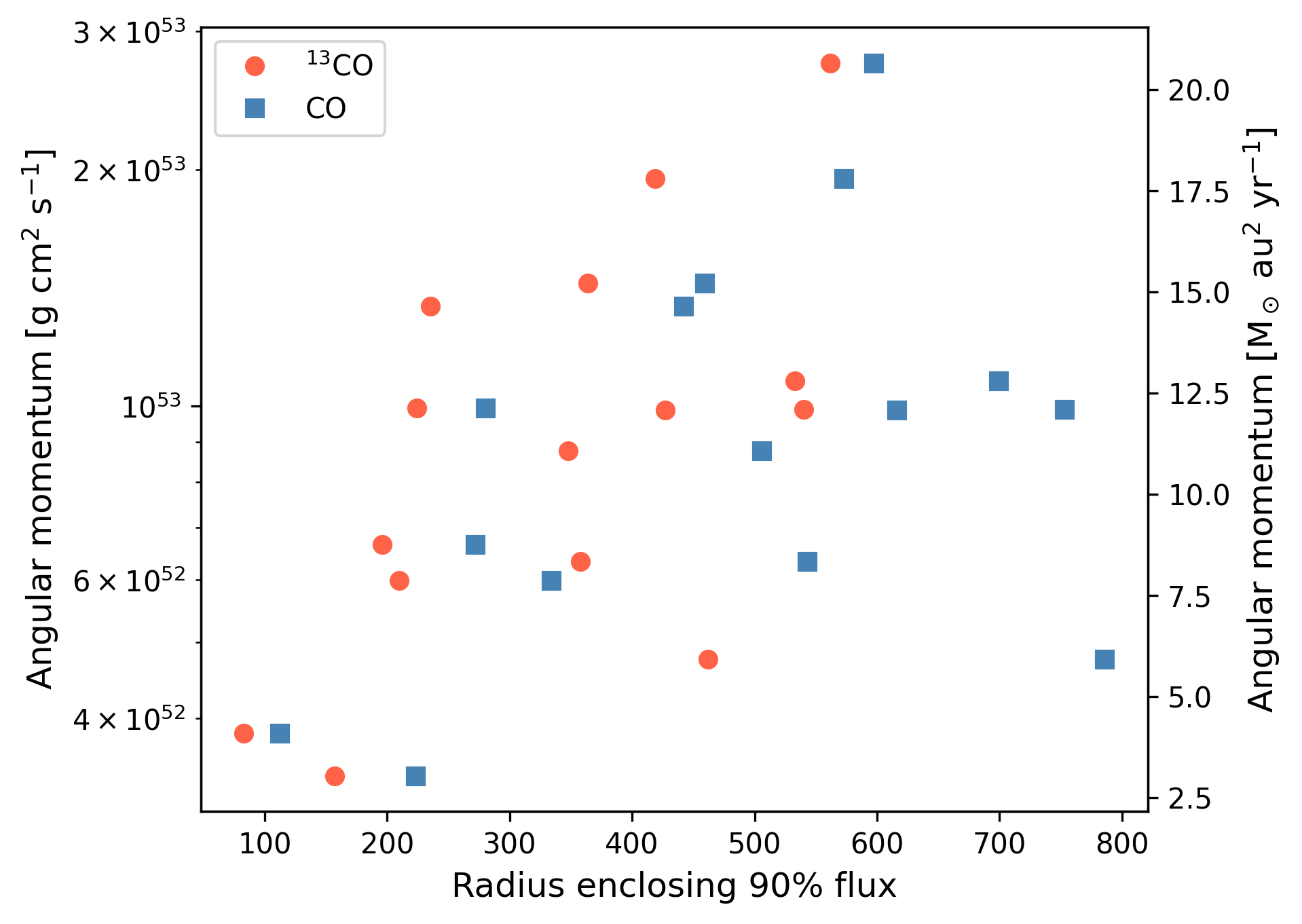}
\caption{Estimated total angular momentum of disks compared to disk radii enclosing 90\% of the $^{13}$CO flux (red circles) and CO flux (blue squares). 
}
\label{fig:AngMomCORadii}
\end{figure}

As mentioned in Section \ref{sec:ang_mom}, the disk radii enclosing 90\% of the $^{13}$CO flux ($R_{\rm disk;\,^{13}CO}$) are significantly correlated with the total angular momentum of disks (Spearman correlation coeff.: 0.59, p-value: 0.02), whereas the correlation is much weaker for the corresponding CO radii ($R_{\rm disk;\,CO}$; Spearman correlation coeff.: 0.36, p-value: 0.19). This is further illustrated in Figure \ref{fig:AngMomCORadii}, where $R_{\rm disk;\,^{13}CO}$ (red circles) scales more strongly with angular momentum than $R_{\rm disk;\,CO}$ (blue squares). We note that the lack of correlation for CO is primarily driven by three disks with somewhat large $R_{\rm disk;\,CO}$ but relatively low angular momentum. These disks, in increasing order of $R_{\rm disk;\,CO}$, are: LkCa 15, IM Lup, and DM Tau.

\section{Age dependence of theoretical infall} \label{app:age}

\begin{figure}[htbp]
    \centering

    \begin{subfigure}{0.4\textwidth}
        \centering
        \includegraphics[width=\linewidth]{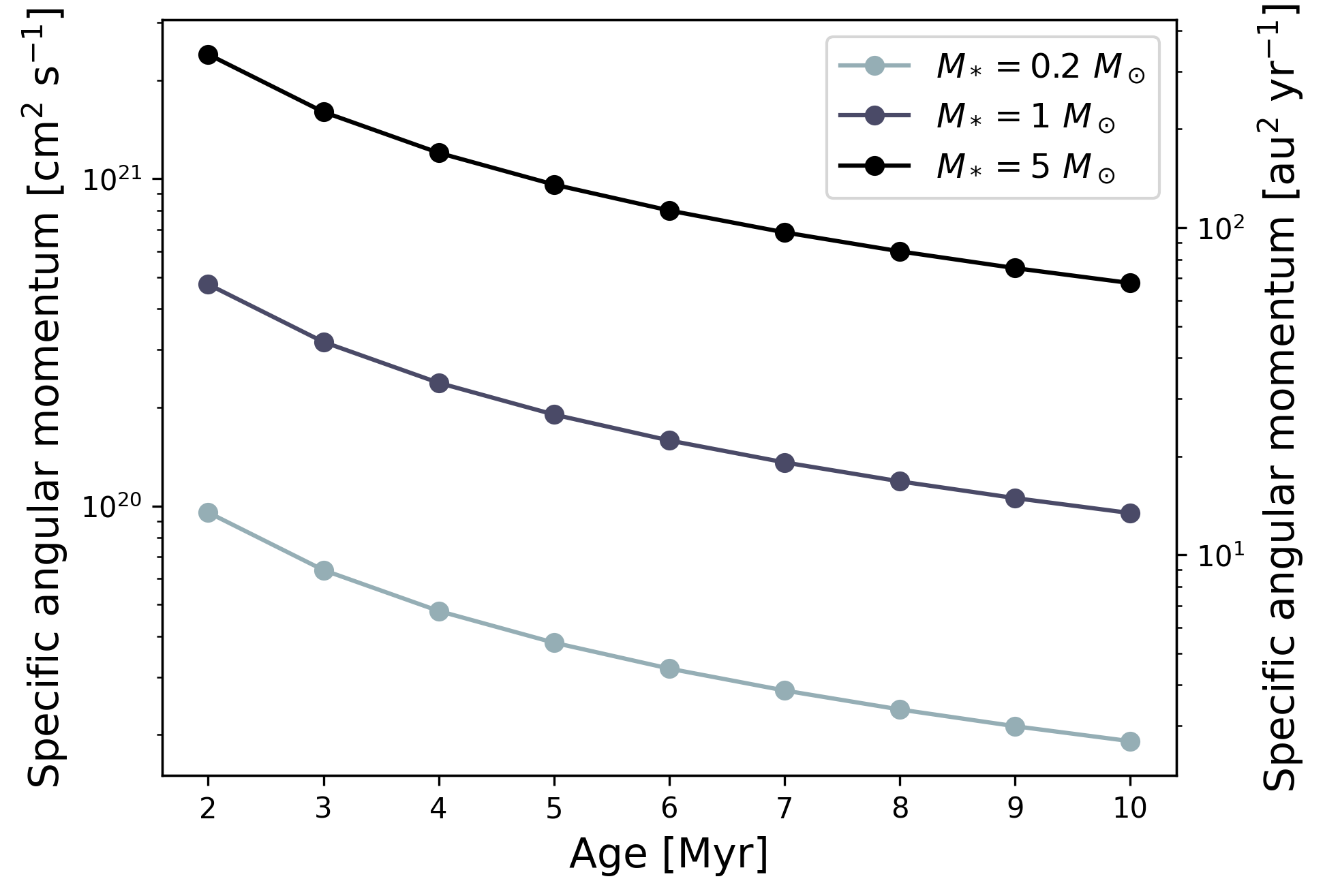}
        \label{fig:SpecificAngMomAge}
    \end{subfigure}

    \vspace{-0.5em}

    \begin{subfigure}{0.4\textwidth}
        \centering
        \includegraphics[width=\linewidth]{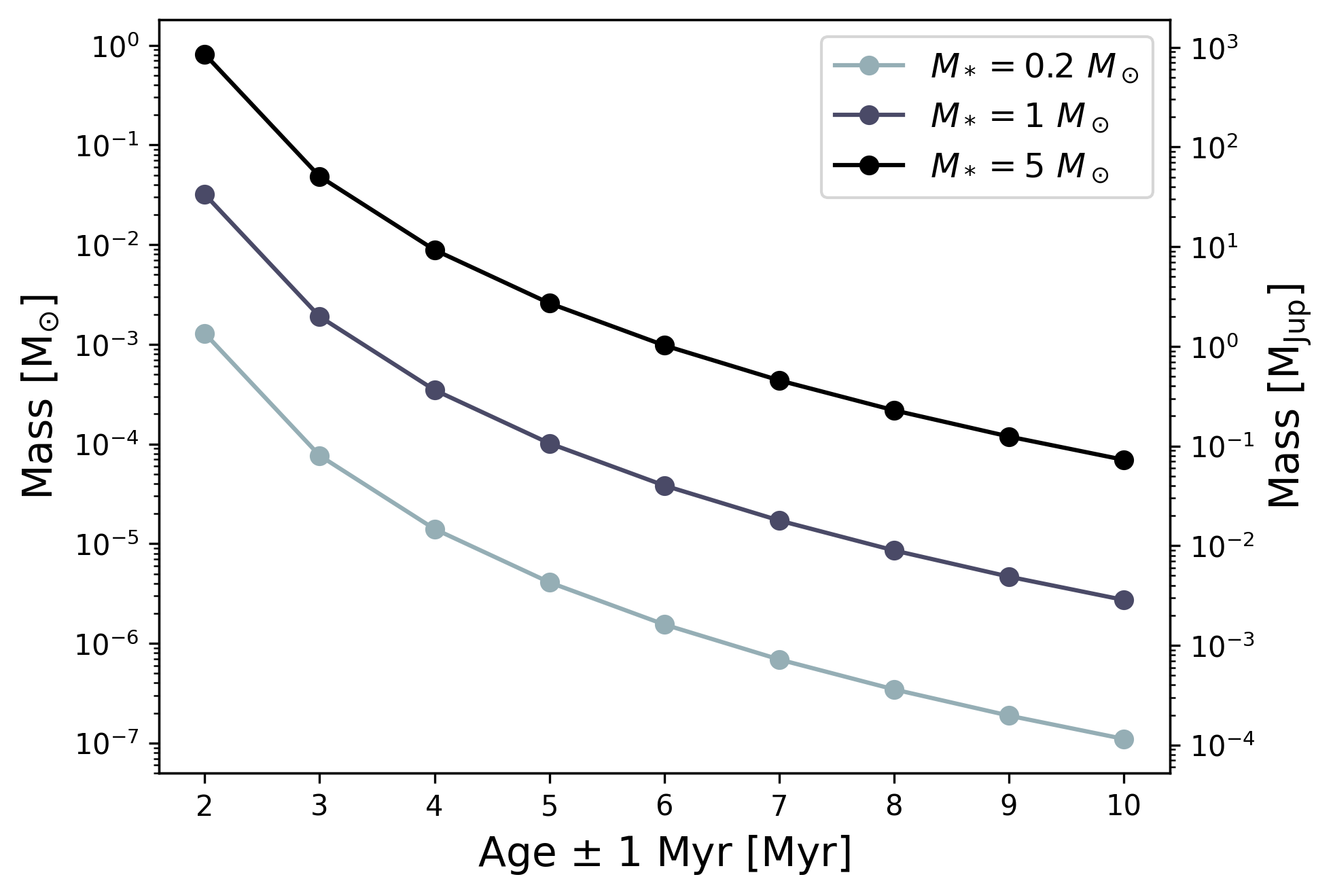}
        \label{fig:MassAge}
    \end{subfigure}

    \vspace{-0.5em}

    \begin{subfigure}{0.4\textwidth}
        \centering
        \includegraphics[width=\linewidth]{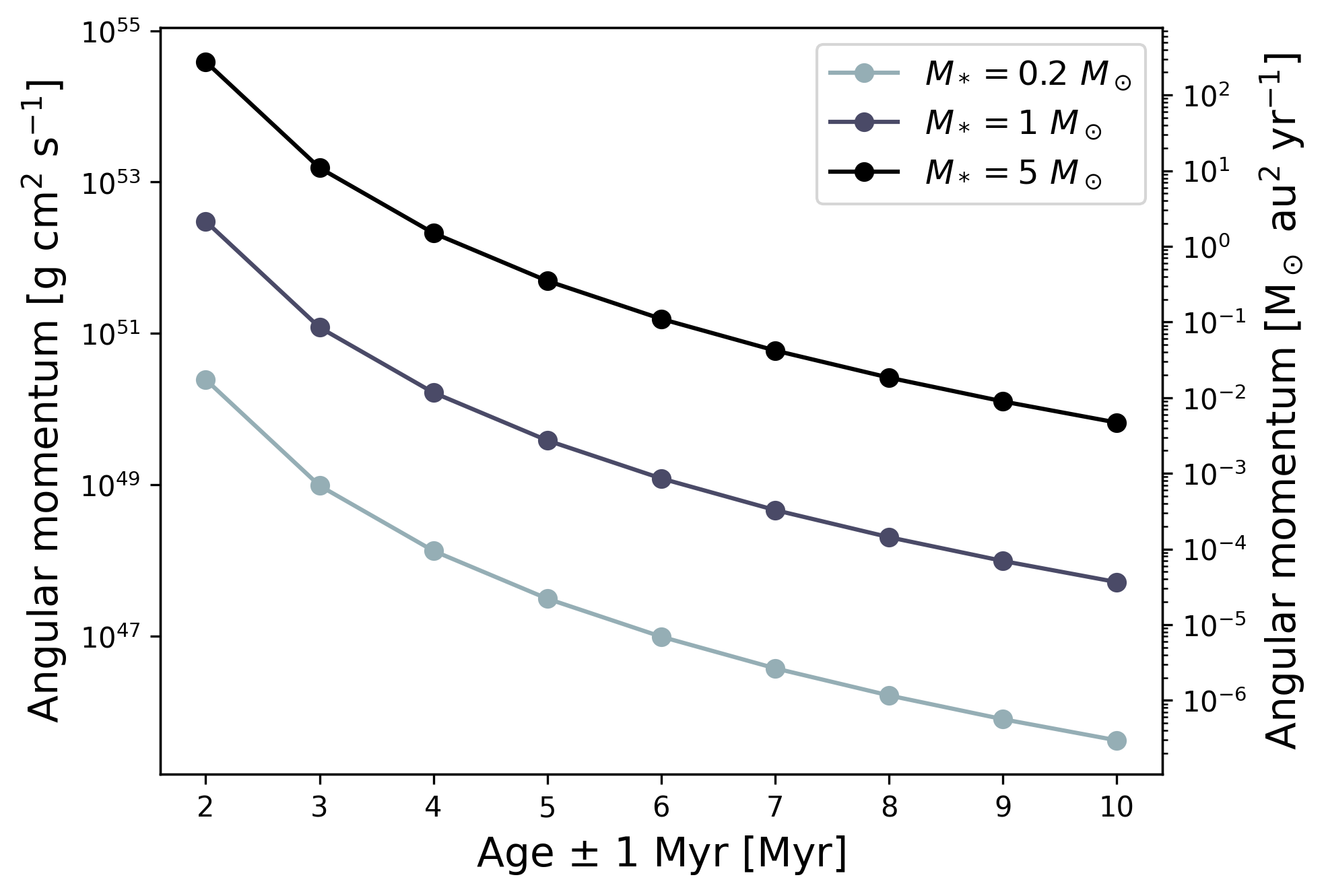}
        \label{fig:AngMomAge}
    \end{subfigure}

    \caption{
    Specific angular momentum (top panel), mass (middle panel), and total angular momentum (lower panel) of infalling material as a function of the age of the cluster, based on the theoretical model presented in \citet{Padoan2025} (see Section \ref{sec:infall_pred} for more details).
    The three lines in each panel represent different stellar masses, as denoted in the legend.
    For the mass and total angular momentum, each value represents the quantity integrated over a $\pm1$~Myr interval around the labeled age; for example, the value at 5~Myr corresponds to infall occurring between 4--6~Myr.
    }
    \label{fig:AgeComparison}
\end{figure}

Figure \ref{fig:AgeComparison} illustrates the age dependence of infalling material as expected from theoretical models of \citet{Padoan2025}, introduced in Section \ref{sec:infall_pred}. Although all the three quantities (specific angular momentum, mass, total angular momentum) decrease with age, the decline is relatively mild for specific angular momentum, which decreases by a factor of five between 2 and 10~Myr. In contrast, the infalling mass, and therefore the total angular momentum, drops by approximately four orders of magnitude over the same period. 
Therefore, these models suggest that the role of late infall in replenishing disk mass and misaligning disks is greatly decreased with the age of the system.
However, these models are intended to capture broad overall trends; the actual infall is expected to be considerably more chaotic, with a high source-to-source scatter in all three quantities, as demonstrated by the numerical simulations of \citet{Pelkonen2025}.

\end{appendix}
\end{document}